\documentclass[twocolumn,twocolappendix]{aastex631}

\usepackage{amsmath}
\shorttitle{CS and H$_2$CS in the HD~163296 Disk}
\shortauthors{Law et al.}
\graphicspath{{./}{figures/}}

\begin{document}

\title{A Multi-line Analysis of the Distribution and Excitation of CS and H$_2$CS in the HD~163296 Disk}

\author[0000-0003-1413-1776]{Charles J.\ Law}
\altaffiliation{NASA Hubble Fellowship Program Sagan Fellow}
\affiliation{Department of Astronomy, University of Virginia, Charlottesville, VA 22904, USA}

\author[0000-0003-1837-3772]{Romane Le Gal}
\affiliation{Université Grenoble Alpes, CNRS, IPAG, F-38000 Grenoble, France}
\affiliation{Institut de Radioastronomie Millim\'{e}trique (IRAM), 300 rue de la piscine, F-38406 Saint-Martin d’Hères, France}

\author[0000-0003-4099-6941]{Yoshihide Yamato}
\affiliation{Department of Astronomy, Graduate School of Science, The University of Tokyo, 7-3-1 Hongo, Bunkyo-ku, Tokyo 113-0033, Japan}

\author[0000-0002-0661-7517]{Ke Zhang}
\affiliation{Department of Astronomy, University of Wisconsin-Madison, 
475 N Charter St, Madison, WI 53706}

\author[0000-0003-4784-3040]{Viviana V. Guzm\'{a}n}
\affiliation{Instituto de Astrof\'isica, Pontificia Universidad Cat\'olica de Chile, Av. Vicu\~na Mackenna 4860, 7820436 Macul, Santiago, Chile}

\author[0009-0009-2320-7243]{Claudio Hernández-Vera}
\affiliation{Instituto de Astrof\'isica, Pontificia Universidad Cat\'olica de Chile, Av. Vicu\~na Mackenna 4860, 7820436 Macul, Santiago, Chile}

\author[0000-0003-2076-8001]{L. Ilsedore Cleeves}
\affiliation{Department of Astronomy, University of Virginia, Charlottesville, VA 22904, USA}

\author[0000-0002-7002-8928]{Greta Guidi}
\affiliation{Institut de Radioastronomie Millim\'{e}trique (IRAM), 300 rue de la piscine, F-38406 Saint-Martin d’Hères, France}

\author[0000-0003-2014-2121]{Alice S. Booth} 
\altaffiliation{Clay Postdoctoral Fellow}
\affiliation{Center for Astrophysics \textbar\, Harvard \& Smithsonian, 60 Garden St., Cambridge, MA 02138, USA}



\begin{abstract}
The abundance and distribution of sulfur-bearing molecules in protoplanetary disks directly influences the composition and potential habitability of nascent planets in addition to providing powerful probes of the physical gas conditions in the disks themselves. Here, we present new and archival ALMA and SMA observations of CS and H$_2$CS, and their C$^{34}$S and H$_2$C$^{34}$S isotopologues, at high-angular resolution (${\approx}$0\farcs2-0\farcs4; 20-40~au) in the HD~163296 disk, which reveal a central cavity and multi-ringed emission structure. These observations comprise the most comprehensive, multi-line CS data in a planet-forming disk to date, spanning a wide range of excitation conditions from E$_{\rm{u}}$=7.1~K to 129.3~K, and include new detections of C$^{34}$S, H$_2$CS, and H$_2$C$^{34}$S in this system. Using these data, we derive spatially-resolved rotational temperature and column density profiles for all species. We find a column density ratio N(H$_2$CS)/N(CS)~$\approx$~0.5, which is comparable to that of the similar MWC~480 disk and suggests that organic sulfur compounds may constitute a large fraction of the volatile sulfur reservoir in disks around Herbig stars generally. We derive $^{32}$S/$^{34}$S ratios of ${\approx}$5 (CS/C$^{34}$S) and ${\approx}$2 (H$_2$CS/H$_2$C$^{34}$S) based on disk-averaged and spatially-resolved analyses. Both values are consistent across these two pairs of optically-thin molecules and are well-below the expected ISM ratio of ${\approx}$22, suggesting significant sulfur fractionation. We also constrain the CS emitting layer ($z/r\lesssim 0.1$) using the vertical separations of the disk surfaces in the channel maps and based on the known 2D gas structure of the HD 163296 disk combined with our excitation analysis.
\end{abstract}
\section{Introduction} \label{sec:intro}

The chemical and physical structure of protoplanetary disks is tightly linked to the properties of nascent planets. Thanks to the high resolution of the Atacama Large Millimeter/submillimeter Array (ALMA), the molecular gas content of disks has now begun to be mapped in fine detail \citep[e.g.,][]{Bergner19,Miotello19, Oberg21_MAPSI,Booth24_HD100546,Rampinelli24}. While many studies have focused on carbon-, oxygen-, and nitrogen-bearing molecules, relatively few have targeted sulfur-bearing species, despite the importance of sulfur for prebiotic chemistry and in determining potential planet habitability \citep{Chen15, Ranjan18}. 

Only a few sulfur-bearing species have been detected in disks. CS is the most commonly-observed molecule \citep[e.g.,][]{Dutrey97, vanderPlas14, Guilloteau16, Semenov18, Legal19, Legal21, Rosotti21, Teague22}, followed by a handful of detections of H$_2$CS, CCS, and H$_2$S \citep{Phuong18, Legal19, Loomis20, Phuong21, Marichalar22, Booth23_HD169142}. Oxygenated species such as SO and SO$_2$ are rare and have been mostly seen in younger disks with signs of active accretion, large-scale structures, and/or planet-induced shocks \citep[e.g.,][]{Fuente10, Pacheco16, Garufi22, Law23_HD16, Booth23, Booth24_IRS48, Huang24, Yoshida24, Dutrey24, Speedie25}, indicating that the bulk of disk gas is, in general, likely highly-reduced. Few of these studies, however, possessed sufficiently high sensitivity and angular resolution while covering multiple lines spanning a range of excitation conditions necessary to constrain gas physical conditions in detail, even in the commonly-observed CS molecule. Such observations are required to capture important spatial variations across disks, e.g., at the location of dust gaps commonly associated with ongoing planet formation. Given its ease of detection in surveys and use as a tracer of disk gas properties \citep[e.g.,][]{Teague16, Teague18_CS}, understanding the detailed distribution and excitation of CS is of critical importance. 

Moreover, the exact nature of the sulfur reservoir in disks remains highly uncertain. Many S-bearing species not yet detected in either the interstellar medium (ISM) or disks are readily observed in Solar System objects, including comets and meteorites \citep[e.g.,][]{Jewitt97, BM04, Biver16, Calmonte16, Paquette17,Altwegg22}. Chemical models suggest that volatile sulfur is significantly depleted in protoplanetary disks \citep{Legal21, Keyte24, Keyte24_HD169142} and much of this missing sulfur may be locked up in the mantles of icy dust grains \citep[e.g.,][]{Millar90, Ruffle99, Laas19, Fuente23} or in refractory components \citep{Kama19}. However, only a small fraction (few percent) of sulfur has been detected in interstellar ices \citep{Boogert15}, while gas-phase H$_2$S has only been detected in two Class~II disks so far \citep{Dutrey11, Phuong18, Marichalar22}, despite it being the main sulfur-carrier in comet 67P/C-G \citep{Calmonte16}. Overall, this means that the distribution of sulfur-containing materials, including the volatile sulfur budget, is still poorly-constrained in disks.

Recent ALMA results, however, have begun to offer additional insights into disk sulfur chemistry \citep[e.g.,][]{Semenov18, Law23_HD16, Keyte23, Keyte24, Booth24_IRS48, Dutrey24}. For instance, \citet{Legal19, Legal21} found an unexpectedly elevated column density ratio of H$_2$CS with respect to CS in the disk around the Herbig star MWC~480, which suggests that a substantial part of this sulfur reservoir may be in the form of organic compounds. Additionally, \citet{Booth24} identified potentially anomalous, non-ISM sulfur isotopic ratios in the IRS~48 disk, which indicates that significant fractionation may be taking place. Taken together, these findings imply that observations of more complex, organic species such as H$_2$CS in addition to simple molecules like CS along with their respective isotopologues provide one promising avenue to constrain this underlying sulfur reservoir, and to address the importance of inheritance versus \textit{in situ} disk processes that may, in turn, alter this reservoir.

In this paper, we present the most comprehensive set of CS lines at high spatial resolution in a protoplanetary disk to date. We also include new detections of  C$^{34}$S, H$_2$CS, and H$_2$C$^{34}$S in the HD~163296 disk. We use these new observations to derive spatially-resolved excitation conditions and sulfur isotopic ratios. In Section \ref{sec:HD163296_disk}, we briefly describe the HD~163296 disk and summarize the ALMA and SMA archival observations used in this work in Section \ref{sec:observations_overview}. We present our results in Section \ref{sec:results} and discuss the origins of the observed emission, including enhanced $^{34}$S isotopic ratios, and compare the sulfur chemistry in HD~163296 to that of other disks in Section \ref{sec:discussion}. We summarize our conclusions in Section \ref{sec:conlcusions}.

\setlength{\tabcolsep}{2.5pt}
\begin{deluxetable*}{lcccccccccccccc}[!]
\tabletypesize{\footnotesize}
\tablecaption{Image Cube and Line Properties\label{tab:image_info}}
\tablehead{
\colhead{Transition} & \colhead{Freq.} & \colhead{Beam}  & \colhead{\texttt{robust}} & \colhead{$\delta$v} & \colhead{RMS} & \colhead{Project} & \colhead{E$_{\rm{u}}$} & \colhead{A$_{\rm{ul}}$} & \colhead{g$_{\rm{u}}$} & \colhead{Int. Flux\tablenotemark{a}}  \\ 
\colhead{} & \colhead{(GHz)} & \colhead{ ($^{\prime \prime} \times ^{\prime \prime}$, $\deg$)} & & \colhead{(km~s$^{-1}$)} & \colhead{(mJy~beam$^{-1}$)} & \colhead{Used} & \colhead{(K)} & \colhead{($\log_{10}$ s$^{-1}$)} & & \colhead{(Jy~km~s$^{-1}$}) }
\startdata
CS J=2--1 &  97.980953 & 0.30$\times$0.30, 0.0 & 0.5,~taper & 0.50 & 0.55 & 2018.1.01055.L & 7.1 & $-$4.776 & 5 & 0.307~$\pm$~0.017 \\
CS J=3--2 & 146.969029 & 0.23$\times$0.18,$-$76.1 & 2.0,~taper & 0.15 & 2.0 & 2017.1.01682.S & 14.1 & $-$4.218 & 7 & 1.195~$\pm$~0.025 \\
CS J=4--3 & 195.954211 & 0.33$\times$0.33, 0.0 & 0.3,~taper & 0.22 & 1.1 & 2021.1.00899.S & 23.5 & $-$3.828 & 9 & 1.779~$\pm$~0.060 \\
CS J=5--4 & 244.935557 & 3.40$\times$2.74, 1.0 & 2.0 & 1.00 & 72.5 & 2020A-S018 & 35.3 & $-$3.527 & 11 & 1.759~$\pm$~0.193 \\
CS J=7--6 & 342.882850 & 0.21$\times$0.21, 88.3 &  0.5, taper & 0.86 & 4.2 & 2016.1.01086.S & 65.8 & $-$3.077 & 15 & ${>}$1.415\tablenotemark{b} \\
CS J=8--7 & 391.846890 & 0.35$\times$0.35, 0.0  & 0.5,~taper & 0.75 & 11.8 & 2015.1.00847.S & 84.6 & $-$2.900 & 17 & 2.602~$\pm$~0.370 \\
CS J=10--9 & 489.750921 & 0.86$\times$0.66, $-$76.0 & 2.0 & 0.15 & 65.8 & 2015.1.01137.S & 129.3 & $-$2.604 & 21 & 1.334~$\pm$~0.165 \\
C$^{34}$S J=5--4 & 241.016089 & 0.52$\times$0.38, 77.4 & 0.5 & 0.31 & 2.4 & 2016.1.00884.S & 34.7 & $-$3.557  & 11  & 0.287~$\pm$~0.030\\
C$^{34}$S J=6--5 & 289.209068 & 0.38$\times$0.26, 69.1 & 0.5 & 0.13 & 1.4 & 2021.1.00535.S & 48.6 & $-$3.318 & 13 & 0.190~$\pm$~0.019  \\
o-H$_2$CS J=6$_{1,5}-$5$_{1,4}$ & 209.200620 & 0.39$\times$0.33, $-$87.9 & 2.0 & 0.18 & 1.7 & 2021.1.00899.S &  48.3 & $-$3.886 & 39 & 0.152~$\pm$~0.038 \\
o-H$_2$C$^{34}$S J=9$_{1,9}$--8$_{1,8}$ & 299.203444 & 0.37$\times$0.25, 69.2 & 0.5 & 1.20 & 0.47 & 2021.1.00535.S & 85.0 & $-$3.402 & 57 & 0.052~$\pm$~0.014 \\ 
\enddata
\tablecomments{The spectroscopic constants for all lines are taken from the CDMS database \citep{Muller01, Muller05, Endres16} with data from \citet{Bogey81, Bogey82, Ahrens99, Kim03, Gottlieb03, Muller19}.}
\tablenotetext{a}{Uncertainties are derived via bootstrapping and do not include the systematic calibration flux uncertainty (${\sim}$10\%). The image properties and flux for CS J=2--1 are taken from \citet{Oberg21_MAPSI}.}
\tablenotetext{b}{Since the CS J=7--6 observations lack short-baseline data, we consider this flux a lower limit.}
\end{deluxetable*} \vspace{-24pt} \setlength{\tabcolsep}{4pt}

\section{The HD~163296 Disk}  \label{sec:HD163296_disk}

HD~163296 is a Herbig Ae star \citep{Fairlamb15} with a stellar mass of 2.0~M$_{\odot}$ \citep{Teague21}, luminosity of 17~L$_{\odot}$, and age of at least 6~Myr \citep{Fairlamb15, Wichittanakom20} located at a distance of 101~pc \citep{Gaia21}. HD~163296 hosts a massive, inclined protoplanetary disk with numerous indications of ongoing planet formation, including multiple dust rings and azimuthal asymmetries \citep{Isella16, Andrews18, Guidi22}, kinematic planetary signatures \citep[e.g.,][]{Teague18_HD163296, Pinte18_HD16, Alarcon22, Calcino22, Izquierdo22}, and meridional gas flows \citep{Teague_19Natur}.

Due to its relative proximity, large size, highly-structured dust, and bright molecular lines, the HD~163296 disk has been extensively studied \citep[e.g.,][]{Thi04, Qi11, Fedele12, Rosenfeld13, Flaherty17, Salinas17, Muro_Arena18, Rich20_HST, Doi21, Calahan21, Oberg21_MAPSI, Booth21_MAPS, Hernandez_Vera24} and has emerged as a benchmark system for understanding gas and dust disk structure in detail. In particular, molecular line observations have detected numerous carbon-, nitrogen-, and oxygen-bearing species, which show a complex, multi-ringed spatial distribution \citep[e.g.,][]{LawMAPSIII}. However, few observations have probed the gas-phase sulfur content of this disk \citep[][]{Legal21, Ma24}. Given the wealth of existing archival data and its well-constrained physical and chemical structure, the HD~163296 disk presents a unique opportunity to comprehensively characterize the distribution and excitation conditions of several sulfur-bearing species (CS, H$_2$CS) and their isotopologues (C$^{34}$S, H$_2$C$^{34}$S) at high-spatial resolution. 

For all subsequent analysis, we adopt the following parameters for the HD~163296 system: PA=133.$^{\circ}$3, incl=46.$^{\circ}$7, v$_{\rm{sys}}$=5.8~km~s$^{-1}$, and M$_*$=2.0~M$_{\odot}$ \citep{Oberg21_MAPSI, Teague21}.

\section{Observations}
\label{sec:observations_overview}

\subsection{Archival Data and Observational Details}
\label{sec:archival_data}

We compiled both published and archival observations of the HD~163296 disk from ALMA and the Submillimeter Array (SMA)\footnote{The Submillimeter Array is a joint project between the Smithsonian Astrophysical Observatory and the Academia Sinica Institute of Astronomy and Astrophysics and is funded by the Smithsonian Institution and the Academia Sinica.}. The ALMA observations span Band 3 to Band 8, and include multiple transitions of CS and C$^{34}$S and a single transition of H$_2$CS and H$_2$C$^{34}$S. Each archival project was initially calibrated by the ALMA staff using the ALMA calibration pipeline and
the required version of CASA \citep{McMullin_etal_2007, CASA22}. For all subsequent analysis, we used CASA \texttt{v6.3.0}. Calibration of the SMA archival data was done using the MIR software package\footnote{\url{https://lweb.cfa.harvard.edu/~cqi/mircook.html}} and followed standard procedures (see Appendix \ref{sec:appendix:observational details} for additional details). Below, we describe the subsequent data reduction, including self-calibration when performed, along with the typical angular and velocity resolution of each of the lines. For programs whose data have already been published, observational details can be found in the references provided. In the case of new and unpublished archival data, we list the full observational details.

\begin{figure*}
\centering
\includegraphics[width=1\linewidth]{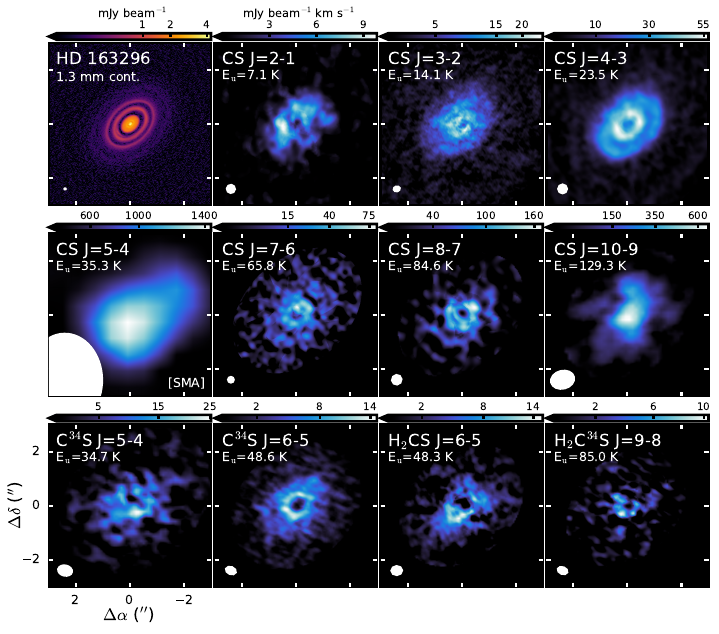}
\caption{1.3~mm continuum image \citep{Andrews18, Huang18_DSHARPII} and zeroth moment maps of multiple CS, C$^{34}$S, H$_2$CS, and H$_2$C$^{34}$S transitions in the HD~163296 disk. Color stretches were individually optimized and applied to each panel to increase the visibility of faint structures. The synthesized beam is shown in the lower left corner of each panel.} 
\label{fig:figure1}
\end{figure*}

\textit{CS J=2--1}: We obtained publicly-available data products from the Molecules with ALMA at Planet-forming Scales (MAPS) ALMA Large Program \citep{Oberg21_MAPSI, Legal21}, which include Band 3 observations of CS J=2--1 at ${\approx}$0\farcs3 with a channel width of 0.5~km~s$^{-1}$. Self-calibration and imaging details are available in \citet{Czekala21}. 

\textit{CS J=3--2}: For the CS J=3--2 line data, we made use of the ALMA program 2017.1.01682.S (PI: G. Guidi), which included both short baseline (SB) and long baseline (LB) data, which covered the CS J=3--2 line in a spectral window with resolution of 35.3~kHz (${\approx}$0.06~km~s$^{-1}$) at ${\approx}$0\farcs1. The continuum data were previously published in \citet{Guidi22}, who describe the observational details. Following standard self-calibration procedures \citep[e.g.,][]{Andrews18}, we first re-aligned all individual execution blocks (EBs) to a common disk center using the \texttt{fixvis} and \texttt{fixplanets} tasks and then performed two rounds of phase-only self-calibration (\texttt{solint}=`900s', `360s') on the SB data. We then concatenated these data with the LB observations before running an additional round of phase-only self-calibration (\texttt{solint}=`inf'). This resulted in an improvement of ${\approx}2\times$ in the peak continuum SNR. We then applied the resulting calibration tables to the aligned line visibilities.

\textit{CS J=4--3, o-H$_2$CS J=6$_{1,5}-$5$_{1,4}$}: We made use of Band 5 observations at ${\approx}$0\farcs3 from the ALMA program 2021.1.00899.S (PI: K. Zhang). The data comprised three EBs observed from 31 July 2022 to 01 August 2022 and used between 41 and 43 antennas for a total on-source integration time of 107~min. Projected baselines ranged from 15.1-1996.7~m and the maximum recoverable scale (MRS) was 3\farcs8 with a typical PWV of 0.6-1.1~mm. One spectral window was centered on the CS J=4--3 line with a resolution of 141.1~kHz (${\approx}$0.2~km~s$^{-1}$). Another spectral window, which was centered on HC$_3$N J=23--22, also serendipitously covered the o-H$_2$CS J=6$_{1,5}-$5$_{1,4}$ line at the same velocity resolution. Before performing self-calibration, all EBs were re-aligned to common center with the the \texttt{fixvis} and \texttt{fixplanets} tasks. Three rounds of phase (\texttt{solint}=`inf', `120s', `60s') and one round of amplitude (\texttt{solint}=`inf') self-calibration were performed, which improved the peak continuum SNR by ${\approx}$4$\times$. The solutions were then applied to the line data.

\textit{CS J=5--4}: We made use of observations from the SMA program 2020A-S018 (PI: R. Le Gal), which were conducted on 10 August 2020 in the compact configuration of the SMA using 8 antennas. These data had an angular resolution of ${\approx}$3$^{\prime \prime}$ and were binned from a native velocity resolution of 140~kHz (${\approx}$0.2~km~s$^{-1}$) to 1.0~km~s$^{-1}$ to improve SNR during subsequent imaging. The SMA data were calibrated with the MIR software package using the bright quasar 3c454.3 as passband calibrator and Uranus as the flux calibrator. Gain calibration was performed with the quasars 1743-038 and nrao530. Once calibrated, the visibilities were then exported to CASA measurement sets for continuum subtraction and imaging. 

\textit{CS J=7--6}: We used Band 7 observations from ALMA program 2016.1.01086.S (PI: A. Isella) at ${\lesssim}$0\farcs1. The continuum data associated with this program were originally published in \citet{Guidi22}, who describe the observational details. The CS J=7--6 line was covered in a broad spectral window with a resolution of 1,128.9~kHz (${\approx}$0.85~km~s$^{-1}$). These observations contain only long-baselines and thus, have a MRS of ${\approx}$0\farcs95. We obtained the image cubes directly from the ALMA archive.

\textit{CS J=8--7}: We used Band 8 observations from ALMA program 2015.1.00847.S (PI: F. Du) at ${\approx}$0\farcs2. These data were previously published in \citet{Bosman21}, but only reported the non-detection of H$_2^{18}$O line emission. The CS J=8--7 line was covered in a broad spectral window at a resolution of 1,128.9~kHz (${\approx}$0.75~km~s$^{-1}$). We performed two rounds of phase (\texttt{solint}=`900s', `360s') and one round of amplitude (\texttt{solint}=`inf') self-calibration after flagged channels containing strong lines, which resulted in an improvement of ${\approx}$600\% in the peak continuum SNR. All solutions were then applied to the unflagged line data.

\textit{CS J=10--9}: We used Band 8 observations from ALMA program 2015.1.01137.S (PI: T. Tsukagoshi) at ${\approx}$0\farcs7. The [C I] $^3$P$_1$--$^3$P$_0$ line data from this program were previously published in \citet{Alarcon22}, who describe the observations in detail. The CS J=10--9 line is covered in a narrow window at 244.1~kHz (${\approx}$0.15~km~s$^{-1}$) resolution. We performed one round of phase-only self-calibration (\texttt{solint}=`inf'), leading to a ${\approx}$350\% improvement in the peak continuum SNR, and then used these solutions for the full line data.

\textit{C$^{34}$S J=5--4}: We used Band 6 observations from ALMA program 2016.1.00884.S (PI: V. Guzm\'{a}n) at ${\approx}$0\farcs5. These data were originally published in \citet{Hernandez_Vera24} but only focused on the H$_2$CO lines. The C$^{34}$S J=5--4 line was also covered in a spectral window with a resolution of 282.2~kHz (${\approx}$0.3~km~s$^{-1}$). We directly obtained the self-calibration, continuum-subtraction measurement sets described in this work from the authors.

\textit{C$^{34}$S J=6--5, o-H$_2$C$^{34}$S J=9$_{1,9}$--8$_{1,8}$}: We used Band 7 observations from ALMA program 2021.1.00535.S (PI: Y. Yamato) at ${\approx}$0\farcs3. The C$^{34}$S J=6--5 line was covered in a narrow spectral window at a resolution of 141.1~kHz (${\approx}$0.13~km~s$^{-1}$) and the o-H$_2$C$^{34}$S J=9$_{1,9}$--8$_{1,8}$ line was serendipitously covered in a broad window at 1,128.9~kHz (${\approx}$1.0~km~s$^{-1}$). Observational details, the self-calibration process, and a description of the original line detections are presented in \citet{Yamato25}, from which we acquired the self-calibration, continuum-subtraction measurement sets.

\subsection{Imaging}
\label{sec:selfcal_imaging}

For those programs which we obtained directly from the ALMA or SMA archives, we first subtracted the continuum using the \texttt{uvcontsub} task in CASA with a first-order polynomial. All lines were then imaged using the \texttt{tclean} task with Briggs weighting and Keplerian masks generated with the \texttt{keplerian\_mask} \citep{rich_teague_2020_4321137} code. The masks were based on the stellar and disk parameters of HD~163296 and have a maximum radius of 4\farcs2, which is matched to that used for the HD~163296 disk in the MAPS Large Program \citep{Czekala21}. We manually chose Briggs \texttt{robust} parameters to prioritize high SNR line detections. In a few cases, we also binned the image cubes in velocity with channel spacings ultimately ranging from 0.13-1.20~km~s$^{-1}$ depending on the line. All images were made using the ‘multi-scale’ deconvolver with pixel scales of [0,5,15,25] and were CLEANed down to a 4$\sigma$ level, where $\sigma$ was the RMS noise measured across five line-free channels of the dirty image. During imaging, we applied \textit{uv}-tapers to several lines to increase the SNR of the resulting images. Table \ref{tab:image_info} summarizes all image properties.

\begin{figure*}[]
\centering
\includegraphics[width=\linewidth]{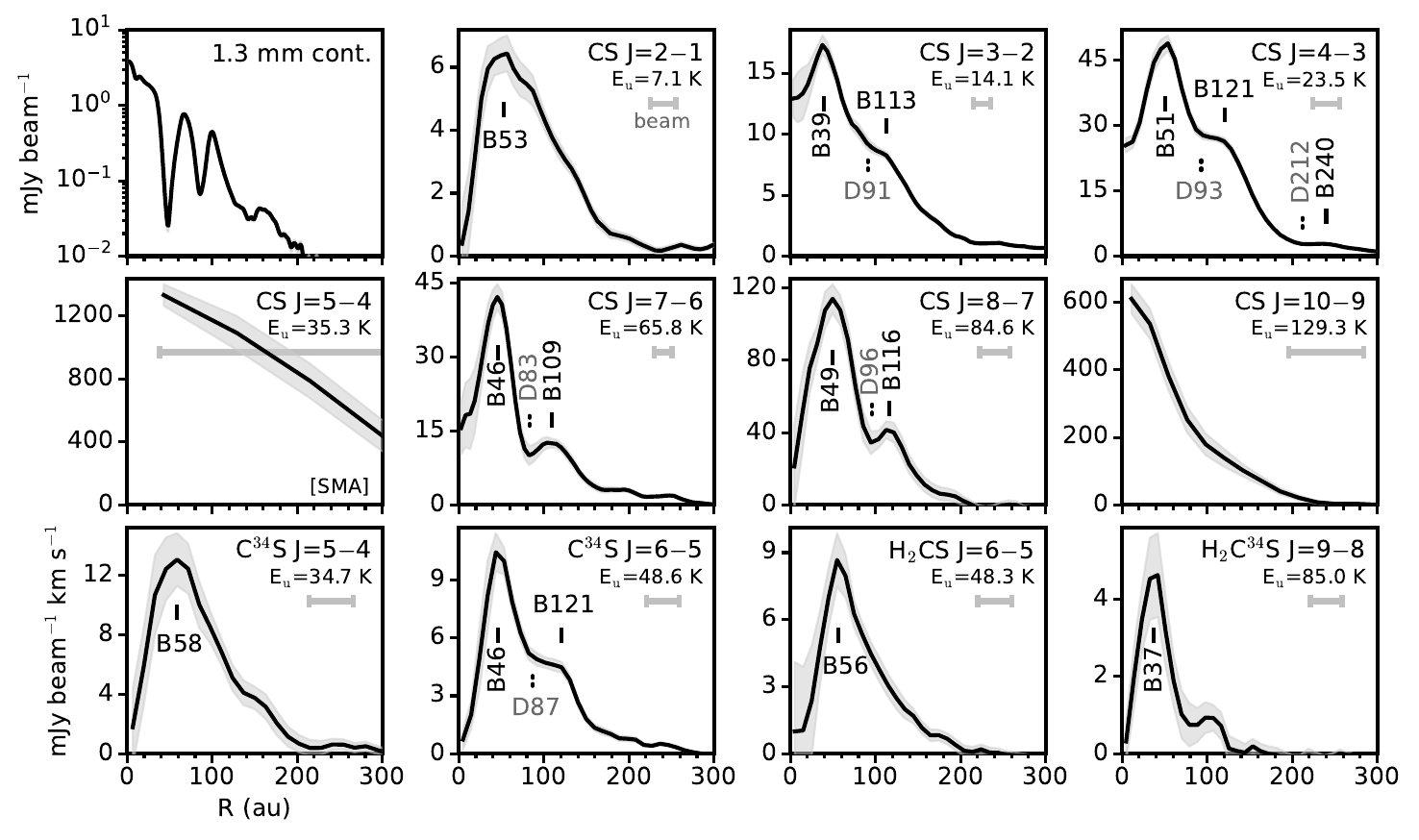}
\caption{Deprojected radial intensity profiles of the 1.3~mm continuum and CS, C$^{34}$S, H$_2$CS, and H$_2$C$^{34}$S lines in the HD~163296 disk. Shaded regions show the 1$\sigma$ uncertainty. Solid lines mark rings and dotted lines denote gaps. The major axis of the synthesized beam is shown in the upper right corner. All profiles are azimuthally-averaged with the exception of C$^{34}$S J=6--5, which was computed in a 45$^{\circ}$ wedge along the disk major axis to better highlight substructure.}
\label{fig:figure2_radial_profiles}
\end{figure*}

\begin{figure*}[]
\centering
\includegraphics[width=0.4\linewidth]{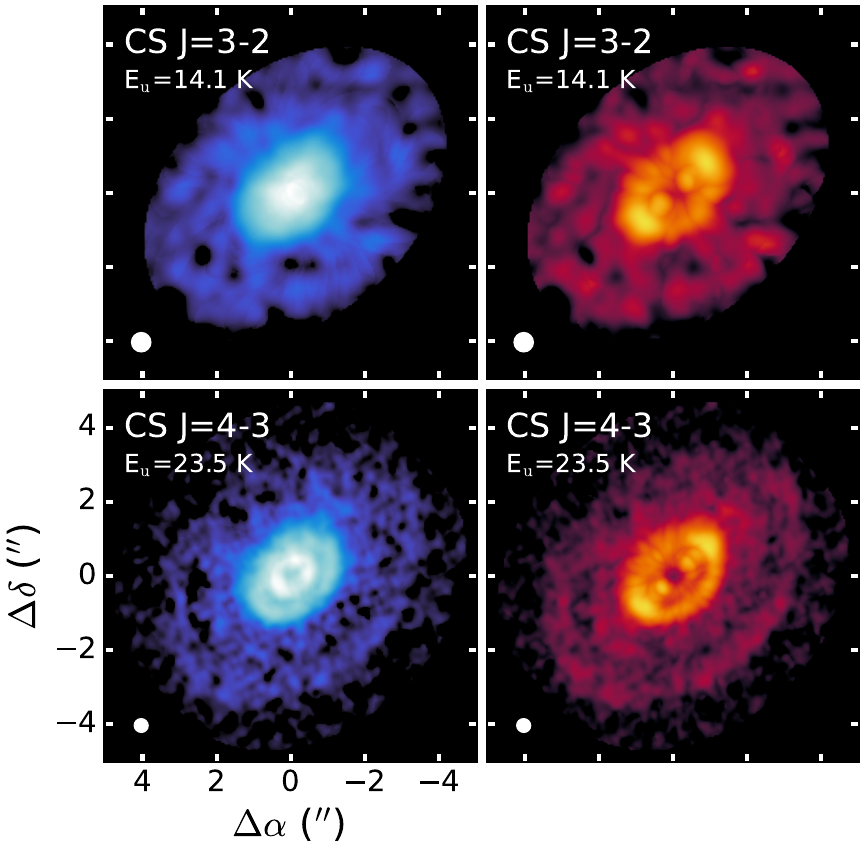}
\includegraphics[width=0.5225\linewidth]{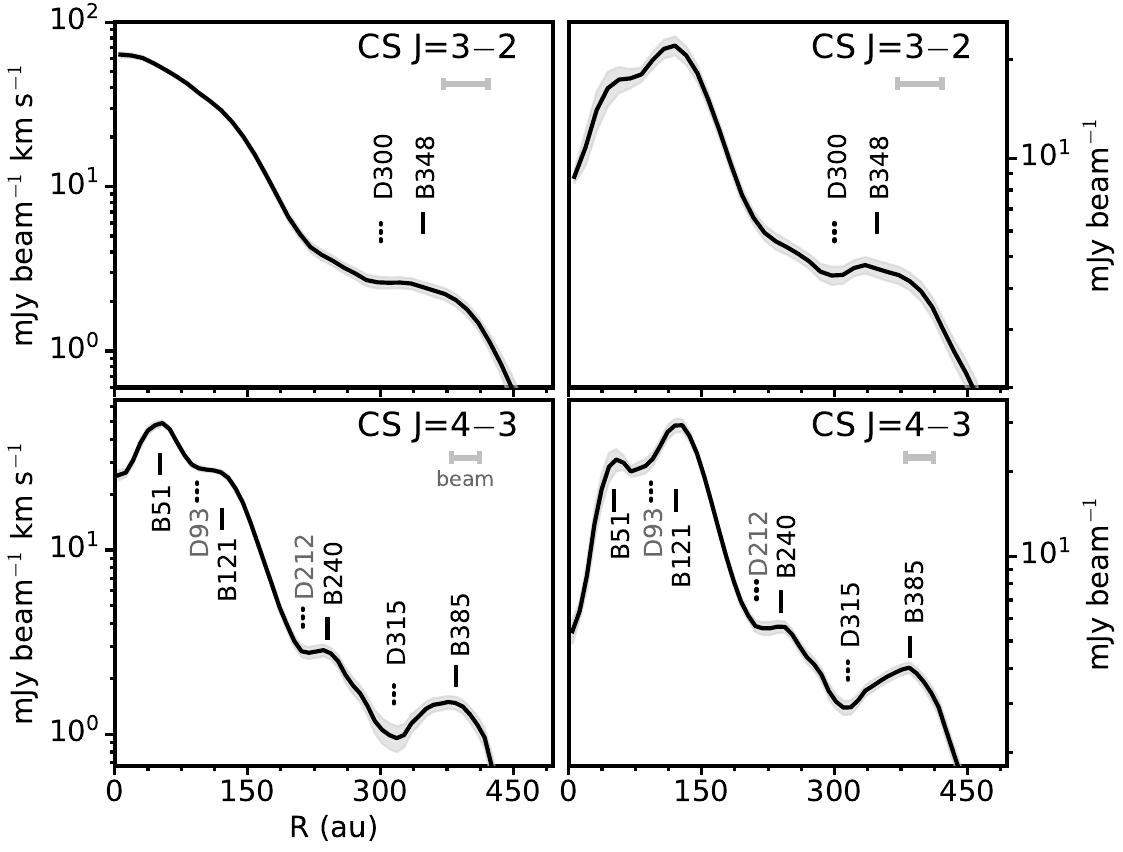}
\caption{Zeroth moment and peak intensity maps and corresponding azimuthally-averaged radial profiles (\textit{columns from left to right}) for CS J=3--2 (\textit{top row}) and CS J=4--3 (\textit{bottom row}). Color stretches have been applied to each of the maps to increase the visibility of outer disk substructure. The CS J=3--2 and J=4--3 data have been smoothed to 0\farcs5 and 0\farcs35, respectively, to highlight the faint, diffuse outer rings.}
\label{fig:figure2_external_rings}
\end{figure*}

\subsection{Moment Maps, Radial Profiles, and Integrated Fluxes}
We generated velocity-integrated intensity, or ``zeroth moment," maps of line emission from the image cubes using \texttt{bettermoments} \citep{Teague18_bettermoments} with no flux threshold for pixel inclusion to ensure accurate flux recovery. All maps were generated using the same Keplerian masks employed during CLEANing. We also produced peak intensity maps using the ‘quadratic’ method of \texttt{bettermoments} (see Appendix \ref{sec:appendix_Fnu_maps}). We computed radial line intensity profiles using the \texttt{radial\_profile} function in the \texttt{GoFish} python package \citep{Teague19JOSS} to deproject the zeroth moment maps. All radial profiles assume a flat, i.e., midplane, emitting surface. We computed line fluxes from spectra extracted using \texttt{GoFish} and the same Keplerian masks used to generate the zeroth moment maps. We estimated uncertainties via a bootstrapping approach as the standard deviation of the integrated fluxes within 500 randomly-generated Keplerian masks spanning only line-free channels. Table \ref{tab:image_info} lists all integrated fluxes. 

\section{Results} \label{sec:results}

\subsection{Spatial Distribution of Emission} \label{sec:line_emission}

Figure \ref{fig:figure1} presents a gallery of zeroth moment maps of all detected lines in the HD~163296 disk and Figure \ref{fig:figure2_radial_profiles} shows the corresponding radial profiles. We characterized line emission substructures using established nomenclature \citep[e.g.,][]{Huang18_DSHARPII, LawMAPSIII}, with local maxima marked with ``B” (for ``bright”) and local minima indicated with “D” (for ``dark”) followed by their radial location rounded to the nearest au. We refer to these features as rings and gaps, respectively.

All lines of CS and H$_2$CS, and their isotopologues, have a ringed structure and central cavity with the exception of CS J=5--4 and CS J=10--9. Given that the radial size of this cavity is  0\farcs4, or ${\approx}$40~au at the distance of HD~163296, the absence of a cavity, and lack of clear rings, in these lines is likely due to their larger beam sizes. In general, CS and C$^{34}$S take the form of two rings at radii of ${\approx}$40-50~au and ${\approx}$110-120~au. The relative contrast of the second ring, i.e., the depth of the gap between the two rings, is a function of the line excitation properties. Lines with low E$_{\rm{u}}$ values only show the inner ring with a broad shoulder (CS J=2--1, C$^{34}$S J=5--4), those with intermediate E$_{\rm{u}}$ values show a plateau-like feature, and those with the highest E$_{\rm{u}}$ values have a well-separated, double-ringed emission morphology (CS J=7--6, J=8--7). This pattern is not an artifact of differing angular resolutions and is clearly seen in the beam-matched profiles (see Section \ref{sec:radially_resolved_Ncol}). 

H$_2$CS instead shows only a single ring at 56~au, which is similar to that of the inner CS ring, and a prominent emission shoulder extending to larger radii, while H$_2$C$^{34}$S peaks in a narrower ring at a smaller radius of 37~au. There is also a possible secondary ring in the H$_2$C$^{34}$S profile at ${\approx}$100~au. If real, it would approximately correspond to the location of the CS and C$^{34}$S outer rings and fall in-line with the trend of increasing ring contrast with higher E$_{\rm{u}}$. However, due to its low SNR, we consider this ring to be tentative and more sensitive observations are needed to robustly confirm this feature. Overall, CS, C$^{34}$S, and H$_2$CS possess broadly similar spatial distributions, including radial locations and widths of features and cavity sizes, while H$_2$C$^{34}$S appears to originate from a more compact ring.

In addition to the two inner rings, we also detected an additional two outer rings in CS J=3--2 and CS J=4--3. Figure \ref{fig:figure2_external_rings} shows zeroth moment and peak intensity maps, along with the corresponding radial profiles, of both lines, which have been smoothed to boost the SNR of the faint, diffuse outer rings. Two external rings at ${\approx}$240~au and ${\approx}$385~au are seen in CS J=4--3, while only the outermost, broad ring is detected in the less sensitive CS J=3--2 data. The radial location of the outermost CS J=3--2 ring differs by a few 10s of au from that of the same feature in J=4--3, but this shift is likely attributable to the additional spatial smoothing required to detect this ring in the J=3--2 data. This multi-ringed emission structure is consistent with that seen in other molecules in the same disk \citep[e.g.,][]{LawMAPSIII, Guzman21_MAPS,Bergner21ApJS}. We return to the origin of these rings in the Discussion.

\subsection{Disk-Integrated Excitation} \label{sec:kinematics}

Next, we aim to determine the disk-averaged column densities and excitation temperatures of CS, C$^{34}$S, H$_2$CS, and H$_2$C$^{34}$S toward the HD~163296 disk. Following \citet{Legal19, Legal21}, we used a rotational diagram analysis \citep{Goldsmith99}. We assume all lines are in local thermal equilibrium (LTE), as in general, we expect the gas densities in the warm molecular layers of the HD~163296 disk \citep[e.g.,][]{Qi11, Zhang21} to be well-above the critical densities of the CS and H$_2$CS lines considered here \citep[${\sim}$10$^4$--10$^6$~cm$^{-3}$, e.g.,][]{Shirley15}.

For optically-thin lines, the disk-integrated flux density $S_{\nu} \Delta v$, as measured for each transition in Table \ref{tab:image_info}, is related to the column density of molecules in their respective upper energy state, N$_{\rm{u}}$ via:

\begin{equation}
    N_u = \dfrac{4 \pi S_{\nu} \Delta v}{A_{ul} \Omega hc},
\end{equation}

\noindent where $\Omega$ is the total solid angle covered by the Keplerian mask from which we measured the flux and A$_{ul}$ is the Einstein coefficient. 

The upper state level population N$_u$ is related to the total column density N$_{\rm{T}}$ and rotational temperature T$_{\rm{rot}}$ by the Boltzmann equation:

\begin{equation}
    \dfrac{N_u}{g_u} = \dfrac{N_{\rm{T}}}{Q(T_{\rm{rot}})} e^{-E_u / k_{\rm{B}} T_{\rm{rot}}},
\end{equation}

\noindent where $g_u$ is the upper state degeneracy, $E_{\rm{u}}$ is the upper state energy, and Q is the molecular partition function. For the diatomic molecules CS and C$^{34}$S, we adopt the following partition function:

\begin{equation}
Q(T) \approx \dfrac{k_B T_{\rm{rot}}}{h B_0} + \dfrac{1}{3}
\end{equation}

with a rotational constant B$_0=$ 24495.56 $\times 10^6$~Hz for CS and B$_0=$ 24103.55 $\times 10^6$~Hz for C$^{34}$S \citep{Muller05}. For H$_2$CS and H$_2$C$^{34}$S, we linearly interpolated values of the partition function from the CDMS catalog, from which we also obtained all spectroscopic line data as listed in Table \ref{tab:image_info}.

The H$_2$CS molecule (and its isotopologue) has two different nuclear spin configurations, i.e., ortho and para corresponding to odd and even values of quantum number K$_{\rm{a}}$, respectively. The H$_2$CS and H$_2$C$^{34}$S lines presented here are ortho lines, which can be used to derive a column density of o-H$_2$CS and o-H$_2$C$^{34}$S, respectively. Thus, to derive total column densities, we adopt the statistical ortho-to-para ratio of 3, which is approximately consistent with the ratio (2.70$^{+0.20}_{-0.19}$) measured from H$_2$CO in the HD~163296 disk \citep{Hernandez_Vera24}.

\begin{figure}[]
\centering
\includegraphics[width=\linewidth]{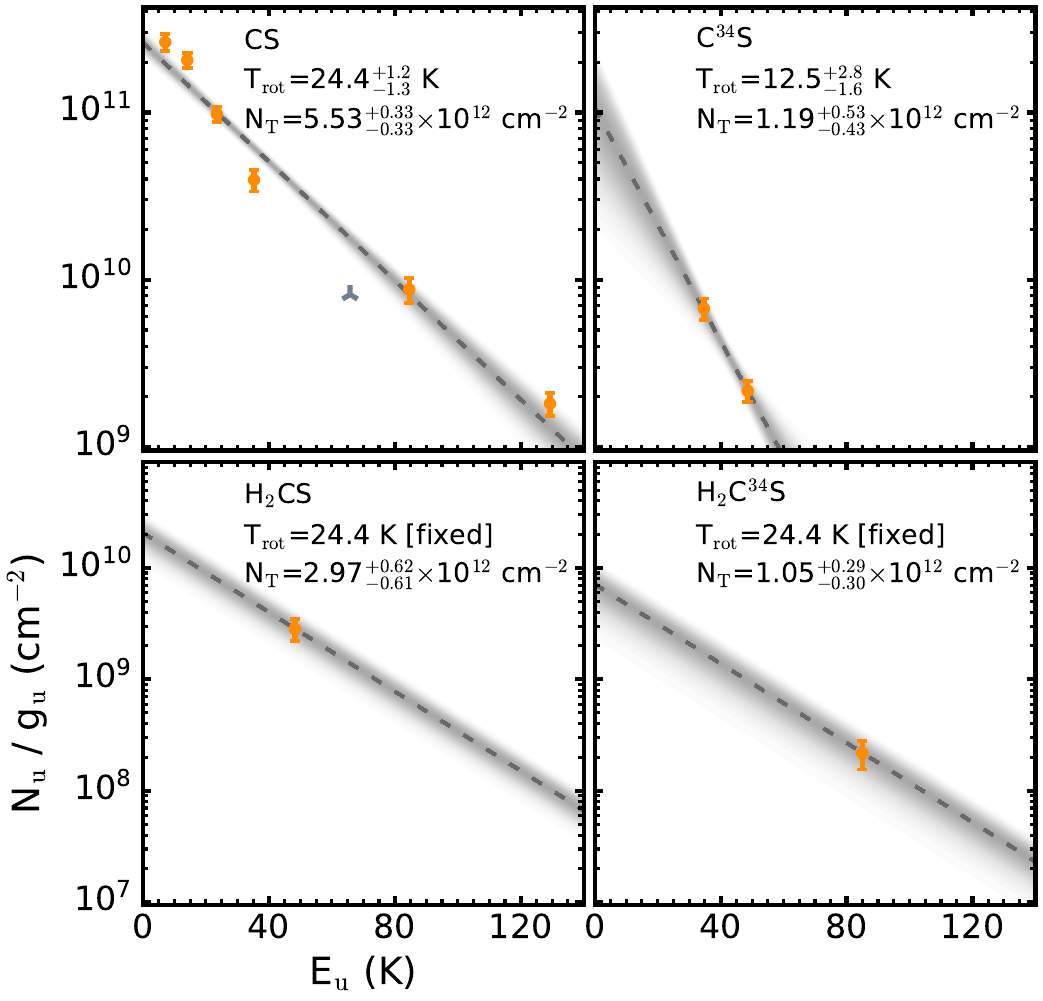}
\caption{Rotational diagrams of CS, C$^{34}$S, H$_2$CS, and H$_2$C$^{34}$S constructed using disk-integrated fluxes of the HD~163296 disk. Gray shaded regions show random draws from the fit posteriors. The lower limit flux value, not used in the fit, for the CS J=7--6 line is shown as a upward, gray triangle. We included a 10\% calibration uncertainty on all measured fluxes.}
\label{fig:figure2}
\end{figure}

Following \citet{Loomis18}, we also included an optical depth correction factor, C$_{\tau}$ = $\tau / 1 - e^{-\tau}$, for the true level populations for each line. We then constructed a likelihood function $\mathcal{L}$(data, N$_{\rm{T}}$, T$_{\rm{rot}}$) for $\chi^2$ minimization and used the affine-invariant Markov Chain Monte Carlo code \texttt{emcee} \citep{Foreman_Mackey13} to generate posterior probability distributions for T$_{\rm{rot}}$ and N$_{\rm{T}}$. For consistency, we adopted the same uniform priors as in \citet{Legal21} of T$_{\rm{rot}} (\rm{K}) = \mathcal{U}(3, 300)$ and N$_{\rm{T}} (\rm{cm}^{-2}) = \mathcal{U}(10^7, 10^{20})$. We used 256 walkers with 2000 burn-in steps and an additional 1000 steps to sample the posterior distribution. We adopt the median as the best-fitting value with uncertainties given by the 16th and 84th percentiles of the posterior distributions, which represent the 1$\sigma$ uncertainties of a Gaussian distribution. We also included a systematic flux calibration uncertainty of 10\% added in quadrature to the statistical uncertainty on all measured line fluxes.

\begin{figure*}[]
\centering
\includegraphics[width=\linewidth]{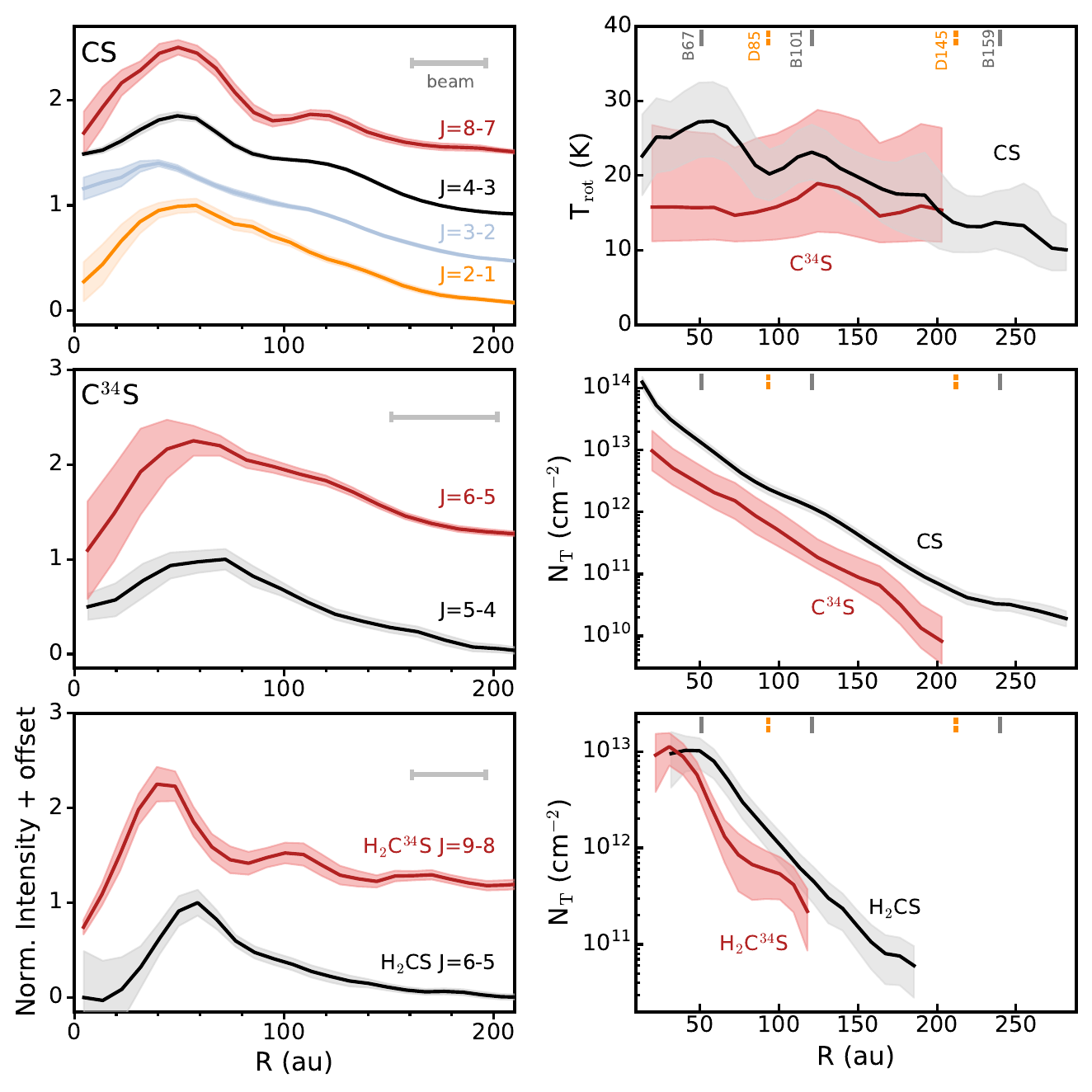}
\caption{Normalized azimuthally-averaged radial intensity profiles (\textit{left column}) of the beam-circularized image cubes used to generate the spatially-resolved T$_{\rm{rot}}$ and N$_{\rm{T}}$ profiles (\textit{right column}). The H$_2$CS and H$_2$C$^{34}$S N$_{\rm{T}}$ profiles were derived assuming the T$_{\rm{rot}}$ profile of CS. Shaded regions show the 1$\sigma$ uncertainty. Orange dashed lines mark continuum gaps and gray solid lines denote continuum rings.}
\label{fig:figure_radially_prof_Ncol}
\end{figure*}

Figure \ref{fig:figure2} shows the fitted rotational diagrams for all molecules. Since only one transition is available for both H$_2$CS and H$_2$C$^{34}$S, we instead fixed the rotational temperature to that derived from CS for both species. This is a reasonable assumption based on the previously-derived H$_2$CS T$_{\rm{rot}}$ in the similar MWC~480 disk \citep{Legal21}. We also experimented with other T$_{\rm{rot}}$ values (from 15~K to 60~K) and found that the derived N$_{\rm{T}}$ values of H$_2$CS and H$_2$C$^{34}$S varied by no more than 2$\times$ and 10$\times$, respectively (see Appendix \ref{sec:appendix:H2CS_trot}). Values of $\tau$ range from 0.0007 (H$_2$C$^{34}$S) to 0.08 (CS), which indicates optically thin emission for all lines and molecules. All fits used thermal line widths, with FWHMs of ${\approx}$1-2~km~s$^{-1}$ measured from spectrally-resolved lines. To provide an upper bound on line optical depths, we adopted the significantly smaller intrinsic line widths of ${\approx}$0.3~km~s$^{-1}$ possible in disks \citep[e.g.,][]{Paneque24}. This resulted in line optical depths that were greater by a factor of ${\approx}$2-4, which implies that a few of the CS lines could, at most, have $\tau\approx$0.2-0.3.

We derive the following disk-integrated values:

\begin{itemize}
    \item T$_{\rm{rot}}$ = 24.4$^{+1.2}_{-1.3}$~K and N$_{\rm{T}}$ = 5.53$^{+0.33}_{-0.33}\times10^{12}$~cm$^{-2}$ for CS
    \item T$_{\rm{rot}}$ = 12.5$^{+2.8}_{-1.6}$~K and N$_{\rm{T}}$ = 1.19$^{+0.53}_{-0.43}\times10^{12}$~cm$^{-2}$ for C$^{34}$S
    \item N$_{\rm{T}}$ = 2.97$^{+0.62}_{-0.61}\times10^{12}$~cm$^{-2}$ for H$_2$CS
    \item N$_{\rm{T}}$ = 1.05$^{+0.29}_{-0.30}\times10^{12}$~cm$^{-2}$ for H$_2$C$^{34}$S
\end{itemize}

We find disk-integrated ratios of: N(CS)/N(C$^{34}$S) =  4.7$^{+3.1}_{-1.6}$; N(H$_2$CS) / N(H$_2$C$^{34}$S) = 2.8$^{+2.0}_{-1.1}$; and N(H$_2$CS) / N(CS) = 0.54$^{+0.15}_{-0.13}$. The derived H$_2$CS/CS ratio is similar to that of the MWC~480 disk \citep{Legal21}, and within a factor of a few of the ratios measured in the HD~100546 \citep{Booth24_HD100546} and HD~169142 \citep{Booth23_HD169142} disks. These high ratios indicate, as previously suggested by \citet{Legal21}, that a substantial part of the volatile sulfur reservoir may be in organic form at least in Herbig sources. In particular, high H$_2$CS/CS ratios are in tension with existing chemical models \citep{Legal19}, which predict H$_2$CS column densities that are 1–2 orders of magnitude below that of CS. As noted in \citet{Legal21}, this underproduction of H$_2$CS is likely due to either missing ice formation pathways or the lack of precise laboratory-measured ice photodissociation rates. Within uncertainties, both CS/C$^{34}$S and H$_2$CS/H$_2$C$^{34}$S ratios indicate a consistent $^{32}$S/$^{34}$S ratio of ${\approx}$2-5, which is at least several times lower than the ISM ratio of ${\approx}$22 \citep{Wilson99}. If we instead adopt the warmer excitation temperature of CS to determine the C$^{34}$S column density, the resulting disk-integrated ratio is N(CS)/N(C$^{34}$S) = 10.4$^{+2.0}_{-1.1}$, which is still considerably lower than the ISM value.

\subsection{Radially-resolved Analysis} \label{sec:radially_resolved_Ncol}

Next, we perform a radially-resolved rotational diagram analysis. First, we selected the subset of lines with sufficiently high angular resolution that resolve one or both inner CS rings, i.e., CS J=2--1, 3--2, 4--3, 8--7, C$^{34}$S J=5--4, 6--5, H$_2$CS J=6--5, and H$_2$C$^{34}$S J=9--8. We excluded the CS J=7--6 line due to its small MRS (see Section \ref{sec:archival_data}), and thus likely missing flux. We re-imaged all CS and H$_2$CS lines with a carefully selected set of \textit{uv}-tapers and \texttt{robust} values to ensure a 0\farcs35 circularized beam, following the approach outlined in \citet{Czekala21}. Due to the larger beam size of the C$^{34}$S J=5--4 observation, we instead tapered both C$^{34}$S lines to a 0\farcs5 beam size. We then used these beam-matched images for the radially-resolved analysis as in the previous Section.

When fitting the C$^{34}$S profiles, we restricted our priors based on the disk-averaged fit and radially-resolved CS values (T$_{\rm{rot}} (\rm{K}) = \mathcal{U}(10, 35)$ and N$_{\rm{T}} (\rm{cm}^{-2}) = \mathcal{U}(10^9, 10^{17})$). To compute the column density profiles of H$_2$CS and H$_2$C$^{34}$S, we adopted the CS T$_{\rm{rot}}$ profile for both species (see Appendix \ref{sec:appendix:H2CS_trot} for more details). As in the disk-averaged analysis, all lines remain optically-thin in the spatially-resolved analysis.

Figure \ref{fig:figure_radially_prof_Ncol} shows the radially-resolved T$_{\rm{rot}}$ and N$_{\rm{T}}$ profiles, along with the beam-matched, azimuthally-averaged radial profiles used to construct them. The T$_{\rm{rot}}$ profile for CS smoothly decreases out to a radius of ${\approx}$250~au, with the exception of a dip (${\approx}$5~K) at ${\approx}$90~au, which is near the location of the D85 dust gap. The C$^{34}$S T$_{\rm{rot}}$ profile is consistent with a constant T$_{\rm{rot}}\approx15$~K out to a radius of 200~au but shows higher uncertainties, which reflect the smaller dynamic range in upper energy of the two lines used in the fitting. 

In terms of N$_{\rm{T}}$, both CS and C$^{34}$S show smoothly declining profiles where N(CS) $>$ N(C$^{34}$S) at all radii. We note a small enhancement in the CS N$_{\rm{T}}$ profile at the location of the second CS ring (${\approx}$120~au). H$_2$CS and H$_2$C$^{34}$S also show generally smoothly declining profiles with the exception of a notable offset in their N$_{\rm{T}}$ profiles, with H$_2$C$^{34}$S peaking ${\approx}$20~au interior to H$_2$CS. This shift reflects the more compact H$_2$C$^{34}$S ring relative to H$_2$CS and results in an overlap region where the column density ratio is on the order of unity. We also observe a flattening of the H$_2$C$^{34}$S profile from ${\approx}$80-120~au, which corresponds to the tentative H$_2$C$^{34}$S secondary ring seen in the intensity profiles (Figure \ref{fig:figure2_radial_profiles}). 

\begin{figure}[]
\centering
\includegraphics[width=\linewidth]{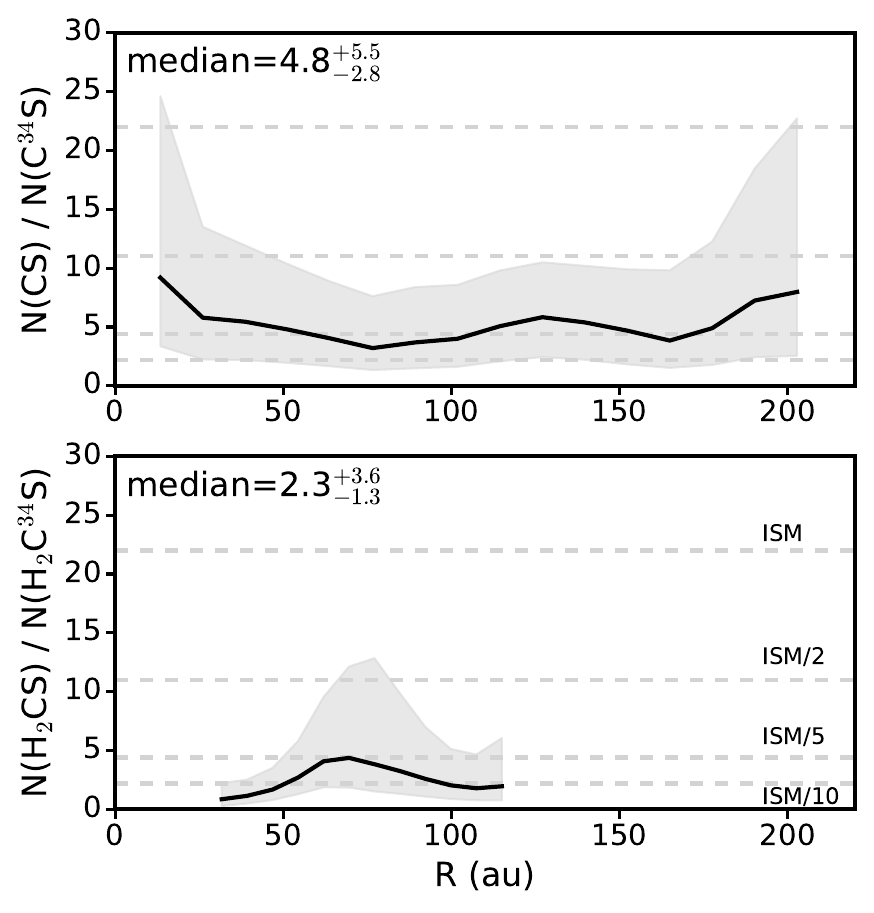}
\caption{Spatially-resolved column density ratios of N(CS) / N(C$^{34}$S) (\textit{top}) and N(H$_2$CS) / N(H$_2$C$^{34}$S) (\textit{bottom}). Shaded regions denote uncertainties and horizontal lines mark fractions of the expected ISM $^{32}$S/$^{34}$S ratio \citep{Wilson99}. Profiles are sampled at one-quarter beam resolution.}
\label{fig:figure_radial_ratios}
\end{figure}

Figure \ref{fig:figure_radial_ratios} shows the spatially-resolved N$_{\rm{T}}$ ratios for both isotopologue pairs of CS/C$^{34}$S and H$_2$CS/H$_2$C$^{34}$S. The spatially-resolved ratios are approximately constant with radius and are broadly consistent with the disk-averaged values. At nearly all radii, the $^{32}$S/$^{34}$S ratios are inconsistent with the ISM value and could be enhanced in $^{34}$S by up to an order of magnitude.

\begin{figure*}[]
\centering
\includegraphics[width=\linewidth]{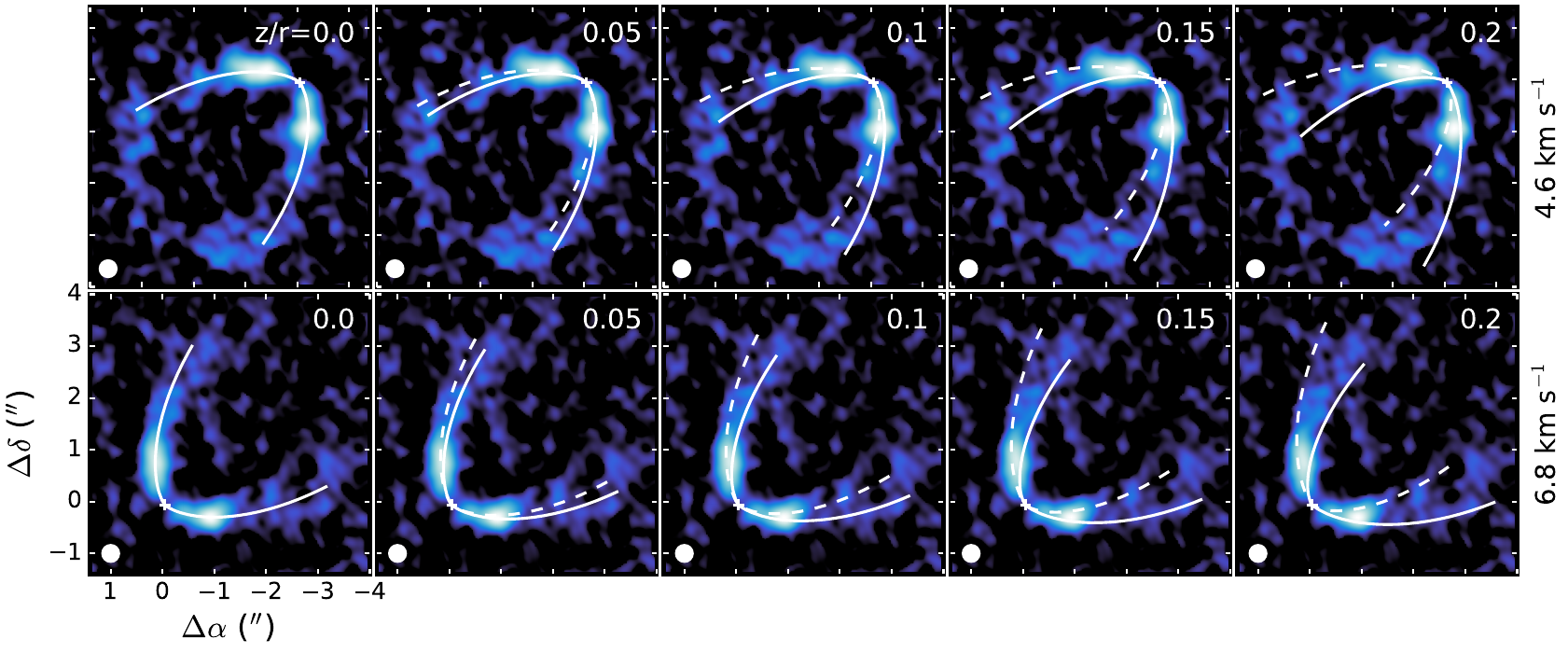}
\caption{Isovelocity contours showing a range of emitting heights (\textit{columns}) for two representative channels of CS J=4--3 at LSRK velocities of 4.6~km~s$^{-1}$ (\textit{top row}) and 6.8~km~s$^{-1}$ (\textit{bottom row}). The solid and dashed lines shows the front and back disk surfaces, respectively. Each tick mark is 1$^{\prime \prime}$. Color stretches have been applied to highlight outer disk structure. The synthesized beam is shown in the lower left corner of each panel.}
\label{fig:figure_emitting_heights_CS_43}
\end{figure*}

\subsection{Emitting Regions of CS and C$^{34}$S} \label{sec:emitting_region_CS}

\subsubsection{Inferred From Channel Maps}

For molecular line emission that arises above the midplane in the warm molecular layer, as is expected for CS and H$_2$CS \citep[][]{Legal19}, there is a spatial separation between the front and back disk surfaces \citep[e.g.,][]{Rosenfeld13, pinte18}. In inclined and vertically-flared disks, such as HD~163296 \citep[e.g.,][]{Teague19Natur, Law21_MAPSIV, Paneque_MAPS}, it is possible to directly resolve this vertical structure at high angular resolution. We searched through the channel maps of all available lines and found that this was possible for CS J=4--3.

We attempted to fit the emitting heights using the channel maps via standard techniques \citep[e.g., \texttt{disksurf}, \texttt{ALFAHOR},][]{disksurf_Teague, Paneque_MAPS}, but the data were of insufficient SNR to enable direct extraction of line emitting heights. However, we can still provide an approximate constraint on the CS emitting surface by overlaying a set of isovelocity contours at different heights and visually inspecting how well they match the observed emission.

In Figure \ref{fig:figure_emitting_heights_CS_43}, we chose a set of representative channels where the separation of disk surfaces is expected to be greatest and labeled emission surfaces from the midplane ($z/r=0$) to a height of $z/r=0.2$ in increments of 0.05. Visually, we can rule out CS originating from altitudes of $z/r \geq 0.2$, which are too elevated relative to the observed emission. While, by-eye, isovelocity contours with heights from $z/r=0.1$ to 0.15 appear to best match the data, we are unable to robustly exclude a midplane or near-midplane origin of CS with the available angular resolution and SNR.

\subsubsection{Inferred from 2D Gas Structure}

In addition to directly resolving the vertical line emission structure, we instead used our radially-resolved T$_{\rm{rot}}$ profile (Figure \ref{fig:figure_radially_prof_Ncol}) to infer the emitting regions of CS and C$^{34}$S based on the known 2D $T(r, z)$ gas temperatures of the HD~163296 disk. To do so, we followed the approach of \citet{Ilee21}. Briefly, we first renormalized the 2D temperature profile of \citet{Law21_MAPSIV}, which was derived from optically-thick CO isotopologues at intermediate-to-high disk altitudes and excluded temperatures below 20~K. This is done to ensure that the CO snowline matches the midplane radius derived in the thermochemical models of \citet{Zhang21}. Then, by assuming that T$_{\rm{rot}}$ is equal to the gas temperature for each molecule, we can determine the emitting region in $(r,z)$-space.

\begin{figure}[]
\centering
\includegraphics[width=\linewidth]{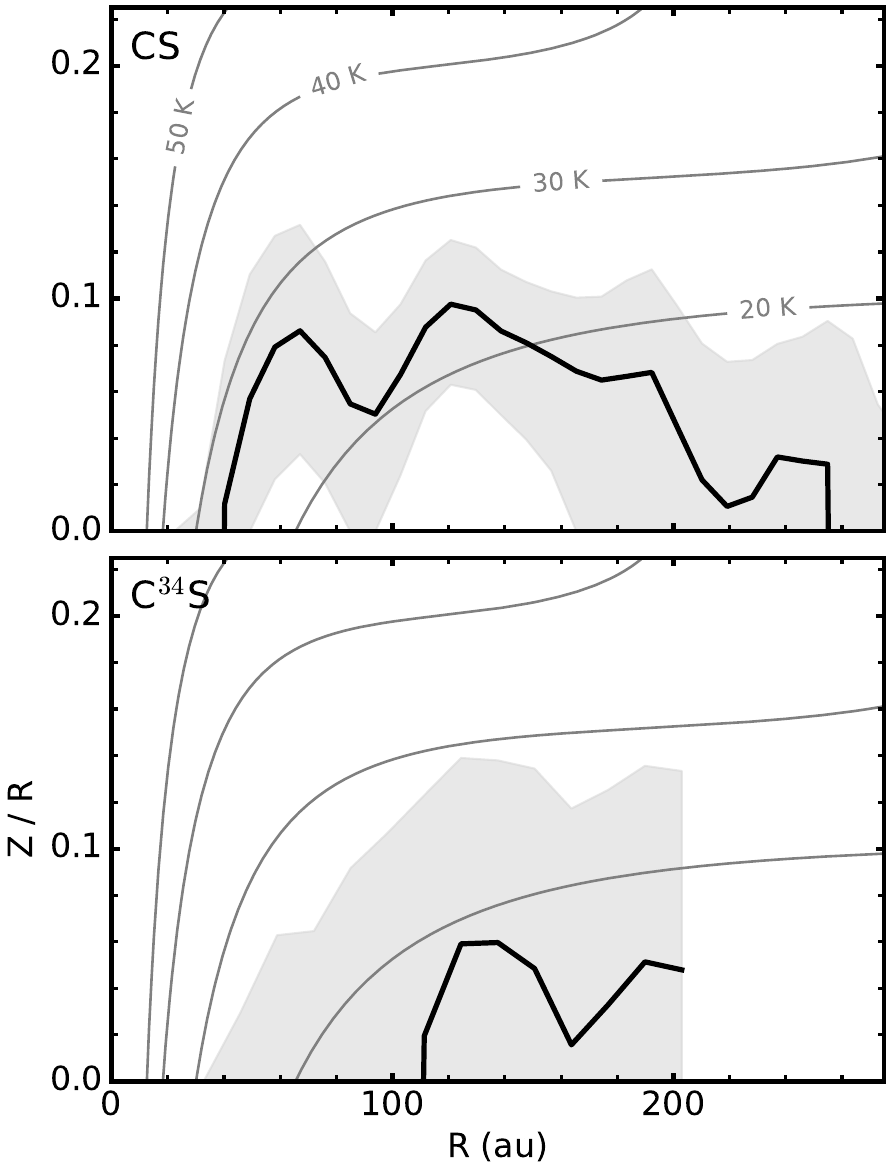}
\caption{Two-dimensional gas temperature structure of the HD~163296 disk (contours) from \citet{Law21_MAPSIV, Ilee21} overlaid with the emitting regions of CS (\textit{top}) and C$^{34}$S (\textit{bottom}) based on their inferred T$_{\rm{rot}}$ profiles.}
\label{fig:figure_zr_heights}
\end{figure}

Figure \ref{fig:figure_zr_heights} shows the inferred emission regions for CS and C$^{34}$S. CS is relatively well-constrained to a height of $z/r\lesssim0.1$, which is consistent with the visual height determination from the CS J=4--3 channel maps in the previous subsection. The apparent decrease in CS emitting heights at a radius of ${\approx}$80-100~au is due to the corresponding dip in the T$_{\rm{rot}}$ profile, which is likely a line excitation effect (see Section \ref{sec:CS_v_continuum}), rather than a true deviation in the emitting heights. The C$^{34}$S emitting region is considerably more uncertain due to the large uncertainties in T$_{\rm{rot}}$.

\section{Discussion} \label{sec:discussion}

\subsection{Dust and Molecular Gas Structure} \label{sec:CS_v_continuum}

\begin{figure}
\centering
\includegraphics[width=1\linewidth]{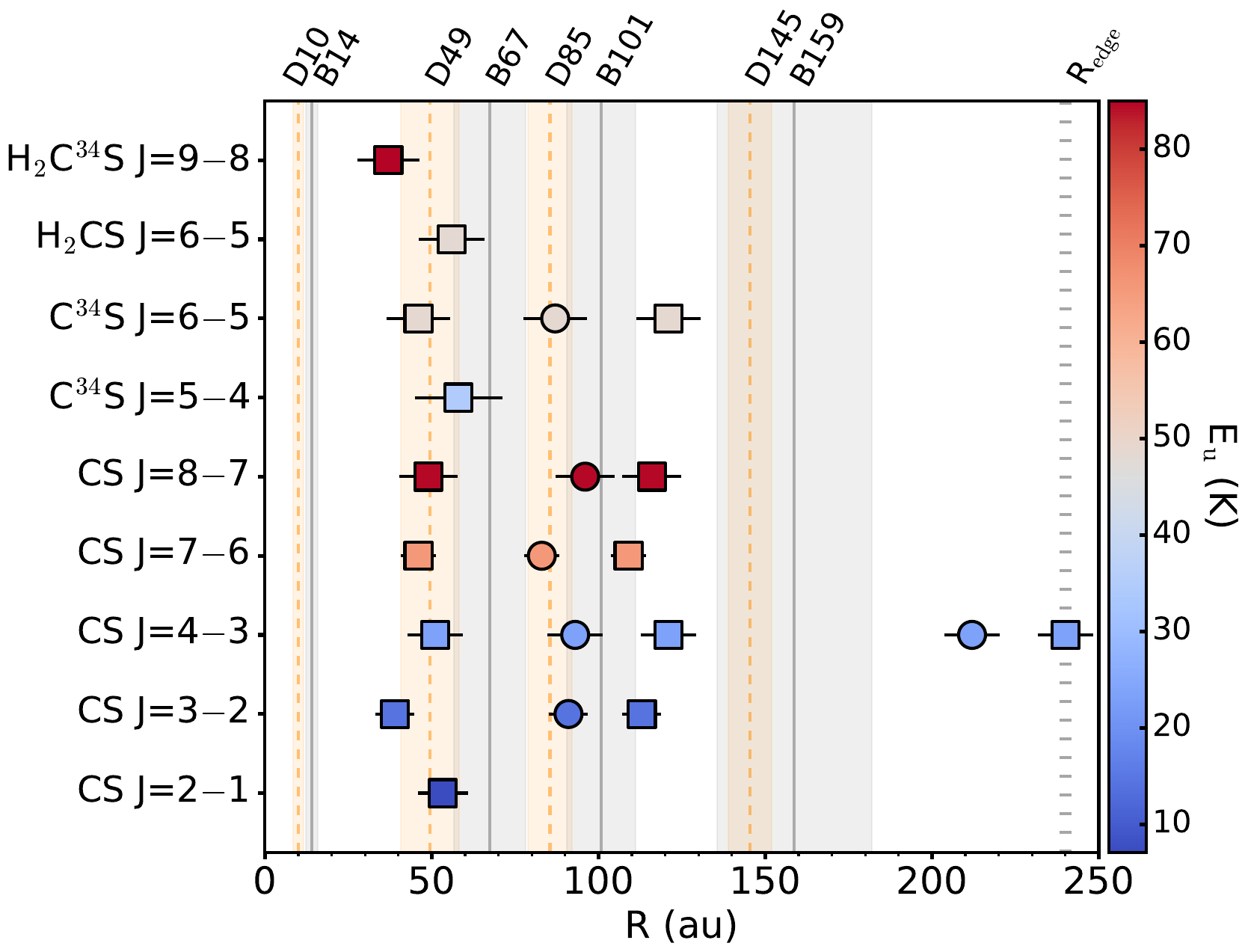}
\caption{Radial locations of CS, C$^{34}$S, H$_2$CS, and H$_2$C$^{34}$S rings (\textit{squares}) and gaps (\textit{circles}) in the HD~163296 disk. Transitions are color-coded according to upper state energy. Orange dashed lines mark continuum gaps and gray solid lines denote continuum rings. The broad dotted line marks the edge of millimeter continuum. We have omitted the outermost CS J=3--2 and J=4--3 rings for visual clarity. Uncertainties are taken to be one-half of the beam major axis.} 
\label{fig:figure_rings_pos}
\end{figure}

Figure \ref{fig:figure_rings_pos} shows the radial locations of all line emission features versus the known annular continuum substructures in the HD~163296 disk \citep{Huang18_DSHARPII,Sierra21_MAPS}. The radial locations of rings and gaps are consistent across different lines. The inner CS and H$_2$CS rings are co-located with the D49 dust gap, while the outer ring is located beyond the dust ring at B101. The gap between these rings is located near the narrow D85 dust gap. One of the outer rings in CS J=4--3 is coincident with the edge of the mm dust. We do not observed any general trends in ring or gap radial locations versus the E$_{\rm{u}}$ of the lines, despite the resolved CS lines spanning nearly 100~K in E$_{\rm{u}}$. 

\begin{figure}[]
\centering
\includegraphics[width=\linewidth]{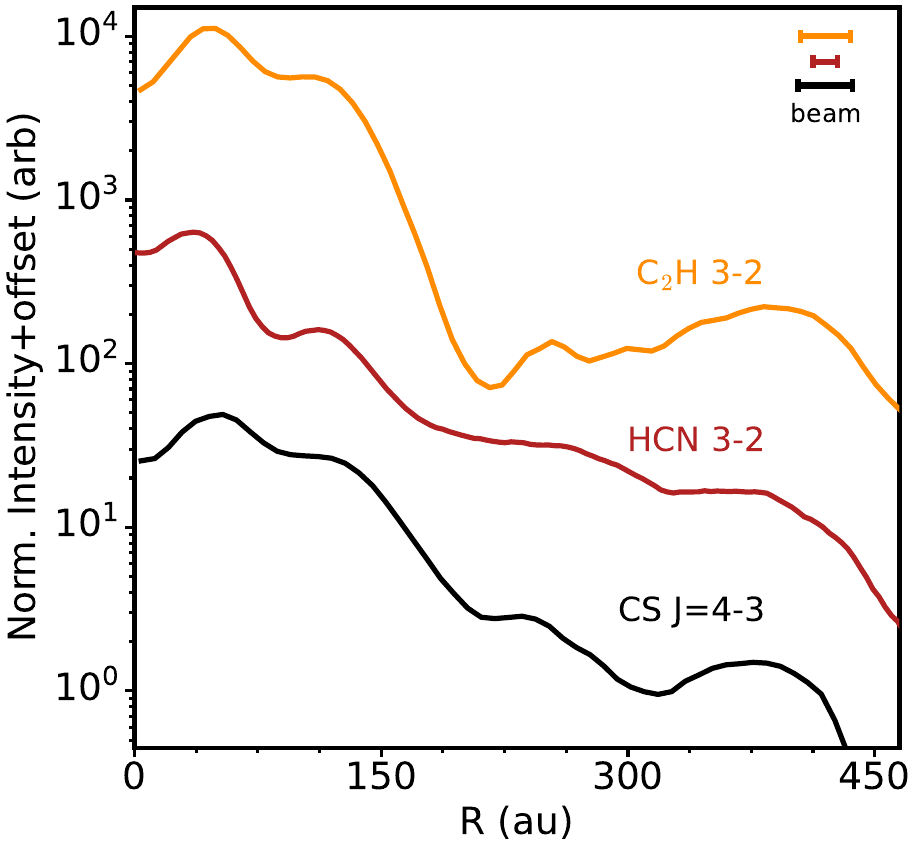}
\caption{Azimuthally-averaged radial intensity profiles of CS J=4--3 compared to C$_2$H 3--2 and HCN 3--2 \citep{Guzman21_MAPS, Bergner21ApJS} in the HD~163296 disk. The four-ringed emission structure is consistent across each of these tracers.}
\label{fig:figure_CS_MAPS}
\end{figure}

The CS emission morphology also largely follows the known gas structure of the HD~163296 disk. As shown in Figure \ref{fig:figure_CS_MAPS}, the radial locations of the four rings seen in the CS J=4--3 line approximately match those of other molecules, such as C$_2$H 3--2 and HCN 3--2 \citep{Guzman21_MAPS, Bergner21ApJS}. All of these lines have comparable excitation properties, e.g., E$_{\rm{u}}$, n$_{\rm{crit}}$, and are expected to originate from similar disk heights of $z/r=0.1$-0.15 \citep[e.g.,][]{Law21_MAPSIV}. Given that each of these molecules have different formation pathways, this indicates that the emission rings are likely tracing the underlying gas density structure in the HD~163296 disk.

In addition to the radial morphology, the tight constraints on the vertical emission structure of CS in HD~163296 provide valuable information on possible formation mechanisms. Although not directly measured, chemical models of the similar disk around the Herbig star MWC~480 from \citet{Legal21} suggest a CS reservoir at altitudes of $z/r\sim0.3$, while models of a generic disk around a Herbig star predict $z/r\gtrsim0.2$ at typical outer disk radii of ${\approx}$100~au \citep{Agundez18}. The relatively lower CS heights in HD~163296 may point to efficient destruction in the disk upper layers, or suggest that significant CS formation is occurring via slower, neutral-neutral reactions rather than the more rapid ion-neutral pathways driven by S$^+$ as predicted by chemical models \citep{Semenov18, Legal19}.

In fact, the CS emitting heights derived here are instead similar to those ($z/r\lesssim0.1$) inferred from high-angular resolution observations of the disk around the T~Tauri star CI~Tau \citep{Rosotti21} and toward the relatively flat disk around the Herbig star HD~100546 \citep{Booth24_HD100546}. However, for both of these sources, the outer disks have evidence for being partially shadowed \citep{Rosotti21, Gravity_CITau, Keyte23}, which would result in cooler temperatures and may inhibit CS formation via photoionization-driven pathways in the upper disk layers. Thus, more general conclusions about CS disk chemistry are limited by the small number of sources with directly mapped CS emission surfaces. High-angular resolution observations toward additional disks, such as those from the ongoing exoALMA Large Program, are vital to determine the typical location of the gas-phase CS reservoirs, and thus, constrain the nature of CS chemistry active in different types of disks.

\citet{Legal21} previously noted the potential presence of azimuthal asymmetries in CS J=2--1 in the HD~163296 disk. Such asymmetries in line emission are relatively rare in disks and are of considerable interest as they often point to dynamical interactions or underlying variations in disk structure \citep[e.g.,][]{vanderMarel13, Temmink23}. Based on the system geometry, we view the disk far side through a known wind \citep{Klaassen13, Booth21_MAPS} and jet \citep{Grady00,Xie21}, which led \citet{Legal21} to speculate about potential local alterations in the CS content due to, e.g., changing C/O ratios or entrained dust reducing line opacity. However, we do not observe any significant asymmetries (Figure \ref{fig:figure1}), and those lines with the highest SNR (i.e., CS J=4--3) appear the most azimuthally uniform. Thus, our multi-line analysis allows us to rule out the presence of any large-scale CS asymmetries in this disk.

\subsection{Local Sub-thermal CS Line Excitation in a Planet-Carved Gap} \label{sec:CS_v_continuum}

One of the most salient morphological features revealed by our multi-line CS observations is the increasing contrast of the inner two inner rings with consecutive J lines. As illustrated in Figure \ref{fig:figure2_radial_profiles}, only a single, wide ring is seen in CS J=2--1, followed by a plateau-like feature in J=3--2 and J=4--3, and ultimately two distinct rings in J=7--6 and J=8--7. The radial location of the gap between these two rings is co-located with the D85 dust gap (Figure \ref{fig:figure_rings_pos}), which was likely carved by an embedded planet \citep[e.g.,][]{Teague18_HD163296,Zhang18}. This gap has been estimated to have a significant local decrease in n$_{\rm{H}}$ \citep[][]{Rab20, Zhang21} and thus, we need to revisit our assumption of LTE and thermalized lines locally at the gap position.

We find that the increasing contrast of the CS rings can be explained by sub-thermal line excitation in the higher J lines. The critical densities range from $n_{\rm{crit}}{\approx}$10$^{4}$~cm$^{-3}$ (J=2--1) to 5$\times10^6$~cm$^{-3}$ (J=8--7) at gas temperatures of ${\approx}$20-30~K \citep[e.g.,][]{Shirley15}. While there is still sufficient gas density in the gap to thermalize the lower J transitions, this is no longer true for the higher J lines. As the n$_{\rm{crit}}$ of the lines increase, it approaches the actual gas density and each subsequent J line is progressively less thermalized than the next. By the time we reach CS J=8--7, which has the highest n$_{\rm{crit}}$ of the resolved lines in our sample, this line is significantly sub-thermally excited. Such a scenario not only naturally explains the apparent gap in the radial intensity profiles (Figure \ref{fig:figure2_radial_profiles}), but also predicts a lower CS rotational temperature at the gap location, as seen in Figure \ref{fig:figure_radially_prof_Ncol}. We confirmed this interpretation via RADEX \citep{vanderTak07} calculations, which show that at the typical CS gas conditions in the gap, T$_{\rm{ex}}$ $\approx$ T$_{\rm{kin}}$ for J=2--1 and J=3--2, while the two differ by more than 20\% in the CS J=7--6 and J=8--7 lines. Thus, at the gap location, by using the sub-thermal CS J=8--7 line fluxes in our excitation analysis, we should expect a ${\approx}$5-6~K underestimate of the kinetic temperature, which is exactly what we observe in the T$_{\rm{rot}}$ profile in Figure \ref{fig:figure_radially_prof_Ncol}.

Given that we have a constraint on what density the CS lines become sub-thermally excited, we can use this to independently estimate the approximate gas density of a planet-carved gap in the HD~163296 disk. This is of particular interest given the wide, nearly two order-of-magnitude range in existing planet mass estimates of 0.07-3~M$_{\rm{Jup}}$ derived in the literature so far \citep[][]{Isella16, Zhang18, Teague18_HD163296, Pinte20}. If we assume that sub-thermal excitation becomes significant when n$_{\rm{H}}$ is within a factor of 10 from the critical density of the CS J=8--7 line, we can then estimate the corresponding gas surface density at this location. We find $\Sigma_{\rm{g}} \sim $ 1~g~cm$^{-2}$, which is consistent with inferences based on gas pressure deviations due to a Jupiter-mass planet \citep{Teague18_HD163296}. Our estimate, while quite uncertain, demonstrates the utility of line excitation as a way to independently measure planet masses. Additional high angular and spectral resolution observations of CS J=8--7 --- and the higher n$_{\rm{crit}}$ line of CS J=10--9, which is mostly unresolved in our data --- would allow for detailed non-LTE modeling and enable significantly improved measurements of the gap gas density, and thus, planet mass.

\subsection{Sulfur Isotopic Ratios}

\subsubsection{Evidence of Sulfur Fractionation in the HD 163296 Disk}

Measurements of isotopic ratios provide valuable constraints on the underlying gas reservoir available to forming planets and serve as diagnostics of disk physical conditions and chemical histories \citep[e.g.,][]{Favre15, Visser18, Cataldi21, Furuya22, Bergin24}. However, sulfur fractionation remains little explored in protoplanetary disks, due in part to the relative scarcity of detections of S-bearing molecules and their isotopologues. It is thus unknown to what degree sulfur fractionation is active in disks, or what fractionation processes may be most important in disk environments. Here, we have measured both a disk-averaged and spatially-resolved $^{32}$S/$^{34}$S ratio from two independent optically-thin isotopologue pairs in the HD~163296 disk. In both cases, we find a consistent $^{32}$S/$^{34}$S ratio of ${\approx}$2-5, which indicates a significant enhancement in $^{34}$S from the expected ISM ratio of ${\approx}$22 \citep{Wilson99}.

In disk models, CS efficiently forms in warm disk layers driven by rapid ion–neutral reactions between S$^+$ and small hydrocarbons, while H$_2$CS forms via a gas-phase neutral–neutral reaction of atomic S and CH$_3$ \citep{Legal19}. Because of these differing formation routes, H$_2$CS is expected to lie deeper in the disk than CS due to the expected vertical distribution of atomic S versus S$^+$. The fact that both molecules show comparable $^{32}$S/$^{34}$S ratios, indicates a wide-spread $^{34}$S enhancement across the disk as both the ionized and atomic sulfur volatile reservoirs must share this enhancement. Moreover, a disk-wide $^{34}$S enhancement is also consistent with the nearly constant $^{32}$S/$^{34}$S ratios out to a radius of 200~au, as seen in Figure \ref{fig:figure_radial_ratios}.

\subsubsection{Comparison to Class II Disks, Comets, and Pre-stellar Cores} \label{sec:comparisons_s}

\begin{figure}[]
\centering
\includegraphics[width=\linewidth]{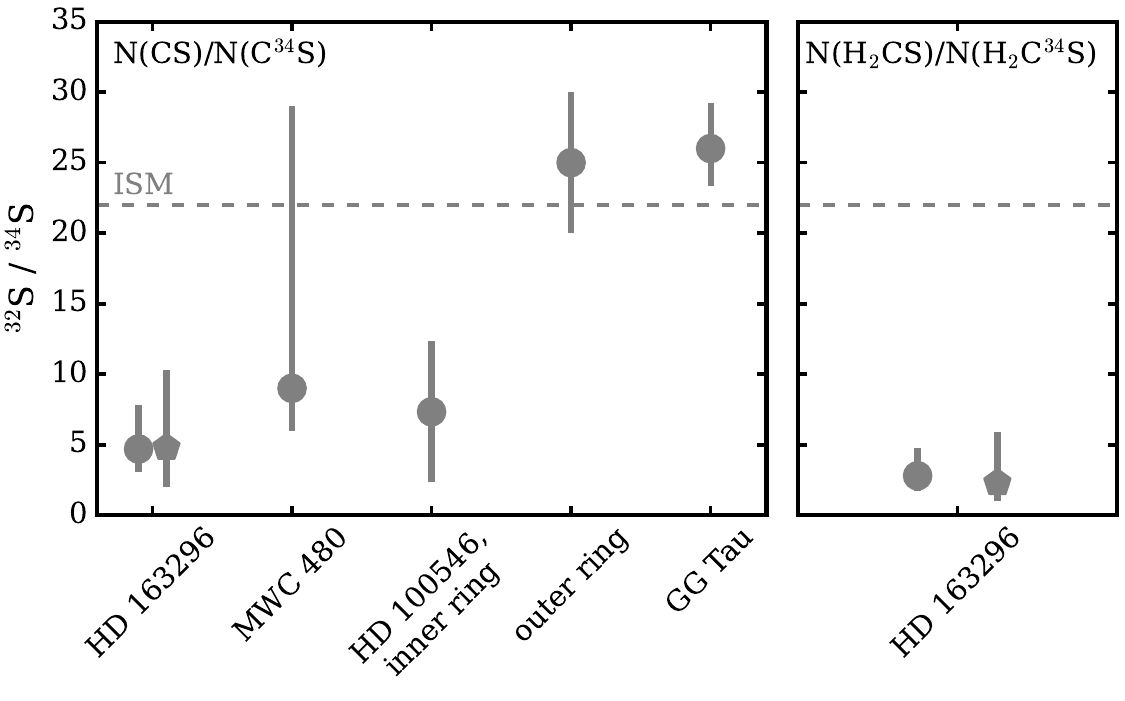}
\caption{Isotopic ratios of $^{32}$S/$^{34}$S derived from column density ratios of CS/C$^{34}$S (\textit{left}) and H$_2$CS/H$_2$C$^{34}$S (\textit{right}). Circle and pentagon symbols are disk-averaged and spatially-resolved measurements, respectively. Literature values are shown for the MWC~480 \citep{Legal21}, GG~Tau \citep{Phuong21}, and HD~100546 \citep{Booth24_HD100546} disks. The ISM ratio is marked as a dashed line \citep{Wilson99}.}
\label{fig:figure_ratios_vs_disks}
\end{figure}

We next aim to compare how the $^{32}$S/$^{34}$S ratio measured in the HD~163296 disk compares to those inferred in other disks, comets, and other ISM settings to shed light on possible fractionation mechanisms. Only a handful of Class~II disks have existing measurements of $^{32}$S/$^{34}$S \citep{Legal21, Phuong21, Booth24_HD100546}, which are shown in Figure \ref{fig:figure_ratios_vs_disks}. The measurements for the MWC~480 and GG~Tau disks represents disk-averaged values, while the ratios in the HD~100546 disk are measured at both the inner (${\approx}$50~au) and outer (${\approx}$200~au) rings. GG~Tau and the outer ring of HD~100546 are consistent with the ISM ratio and show no evidence of enhanced $^{34}$S. However, the inner ring of HD~100546 and MWC~480 appear similar to that of HD~163296, but both of these measurements have associated caveats. As \citet{Booth24_HD100546} note, the measured N(CS)/N(C$^{34}$S) ratio in the inner HD~100546 ring may indicate optically-thick CS rather than an anomalous $^{32}$S/$^{34}$S ratio. On the other hand, while CS in the MWC~480 disk is optically-thin \citep{Legal21}, the large upper range in uncertainty means that we cannot rule out an ISM-like $^{32}$S/$^{34}$S ratio.

We emphasize that only HD~163296 and MWC~480 have isotopic measurements derived from multi-line analysis in both CS and C$^{34}$S, where it was possible to confirm that both tracers are optically thin, while the ratios in GG~Tau and HD~100546 are based on a single line of CS and C$^{34}$S each. If CS is indeed optically-thin in the inner ring of HD~100546, this then suggests that each of the Herbig sources observed so far have an enhanced $^{34}$S ratio. Unlike the HD~163296 disk, which shows a radially-constant $^{34}$S enhancement, this would imply a radial gradient in the $^{32}$S/$^{34}$S ratio in the HD~100546 disk. Taken together, this may point to fractionation pathways that depend on disk properties, e.g., gas temperatures, radiation field, which vary across Herbig versus T~Tauri sources. We also note that, recently, \citet{Booth24} found that both $^{33}$S and $^{34}$S were depleted in the disk around Herbig star IRS~48 as measured via SO isotopologues, while $^{34}$SO$_2$ showed a potential enhancement. However, direct comparisons with our findings are challenging as these inferences of non-ISM-like isotopic ratios were highly dependent on the assumed excitation temperatures. Moreover, the observed volatile sulfur reservoir in IRS~48, unlike most Class~II disks,  is primarily in the form of SO with only a negligible contribution from CS to the total sulfur abundance \citep{Booth24_IRS48}. Nonetheless, these results demonstrate that not all disks show ISM-like sulfur isotopic ratios and illustrate the urgent need for additional multi-line observations of sulfur-bearing isotopologues to assess how common fractionation is in disk settings. 

We can also compare the measured $^{32}$S/$^{34}$S ratio in the HD~163296 disk to those derived in comets and earlier ISM phases, to assess if inheritance or \textit{in situ} processes set this ratio in disks. Isotopic ratios have been thoroughly characterized in comets and no similar $^{34}$S enhancement is found in cometary material measured in a variety of sulfur tracers, including gas-phase molecules and refactory dust \citep[e.g.,][]{Heck12, Biver16, Calmonte17, Paquette17}, which are instead consistent with the local ISM ratio. In contrast, few sulfur isotopic ratios have been robustly measured in earlier ISM phases. Such measurements are particularly difficult in the protostellar phase, since sulfur-bearing molecules often trace outflows and/or the surrounding envelopes, rather the disks themselves \citep[e.g.,][]{Villarmois23}. However, the pre-stellar core L1544 has had its sulfur inventory categorized in detail by \citet{Vastel18}. While CS emission is optically thick in L1544, \citet{Vastel18} report an ISM-like $^{32}$S/$^{34}$S ratio measured from the optically thin H$_2$CS/H$_2$C$^{34}$S column density ratio. Thus, the fractionation taking place in the HD~163296 disk appears distinct both from cometary values and those in low-mass pre-stellar cores, which again points to a potential origin in the differing gas environment present around a higher-mass star.

\subsubsection{Spatial Offset of H$_2$C$^{34}$S Ring}

In Section \ref{sec:line_emission}, we found that the H$_2$C$^{34}$S ring at 37~au appears spatially offset from that of H$_2$CS at 56~au, which corresponds to an offset of approximately half of the synthesized beam size. Similar offsets have been observed in the DCN, HC$^{15}$N, and H$^{13}$CN isotopologues in the PDS~70 disk, which were attributed to active \textit{in-situ} fractionation pathways occurring at the inner cavity wall \citep{Rampinelli24}. If so, we may be seeing a similar effect in the HD~163296 disk in sulfur isotopologues.

Alternatively, given the large difference in E$_{\rm{u}}$ between the H$_2$CS J=6--5 (E$_{\rm{u}}$=48.3~K) and H$_2$C$^{34}$S J=9--8 (E$_{\rm{u}}$=85.0~K) lines, this could instead be an excitation effect. However, we do not see any shifts in the CS or C$^{34}$S inner ring positions (Figure \ref{fig:figure_rings_pos}) as a function of E$_{\rm{u}}$, despite highly-resolved observations spanning a wider range in E$_{\rm{u}}$ from 7.1~K to 84.6~K. The origin and robustness of this shift in H$_2$C$^{34}$S, while intriguing, cannot be definitely confirmed without comparable observations of the same line in both H$_2$CS and H$_2$C$^{34}$S. Such observations will also allow us to robustly determine if this offset is an excitation effect or if it indicates that a different fractionation process is happening in H$_2$C$^{34}$S.

\section{Conclusions} \label{sec:conlcusions}

We presented new and archival ALMA and SMA observations of CS and H$_2$CS, and their C$^{34}$S and H$_2$C$^{34}$S isotopologues, at high-angular resolution (${\approx}$0\farcs2-\farcs0.4; 20-40~au) in the HD~163296 disk. These observations comprise the most comprehensive, multi-line CS data in a planet-forming disk to date, spanning a wide range of excitation conditions from E$_{\rm{u}}$=7.1~K to 129.3~K, and include new detections of C$^{34}$S, H$_2$CS, and H$_2$C$^{34}$S  in this system. We conclude the following:

\begin{enumerate}
    \item Both CS and H$_2$CS, and their isotopologues, show a multi-ringed structure and central cavity. The radial location of the line emission rings and gaps do not show a consistent trend with the mm dust continuum. We also report the discovery of two outer (${>}$200~au) rings in CS J=3--2 and J=4--3. Overall, the sulfur chemistry appears to broadly follow the known molecular gas distribution in the HD~163296 disk.
    \item We derived both disk-averaged and radially-resolved rotational temperature and column density profiles for CS, C$^{34}$S, H$_2$CS, and H$_2$C$^{34}$S. We find that all molecules are optically thin.
    \item We report a N(H$_2$CS)/N(CS) ratio of $\approx$0.5, which is similar to that seen in the Herbig disk MWC~480 and suggests that organic sulfur compounds may constitute a large fraction of the total sulfur reservoir in disks around Herbig stars generally.
    \item The higher contrast of the two inner CS rings in subsequent J lines can be explained by sub-thermal line excitation of the higher J lines. Thus, we provide an independent estimate on the approximate gas density of the planet-carved gap at 85~au in the HD~163296 disk. We find $\Sigma_{\rm{g}} \sim 1$~g~cm$^{-2}$, which is consistent with previous estimates indicating the presence of a Jupiter-mass planet.
    \item We derive robust $^{32}$S/$^{34}$S ratios of ${\approx}$5 (CS/C$^{34}$S) and ${\approx}$2 (H$_2$CS/H$_2$C$^{34}$S) based on both disk-averaged and spatially-resolved analyses. Both values are consistent across two independent pairs of optically-thin molecular traces. This ratio is well-below the expected ISM ratio, which suggests the presence of significant sulfur fractionation in the HD~163296 disk. 
    \item We constrain the CS emitting height to be $z/r<0.2$ based on vertical separations of the disk surfaces in the CS J=4--3 channel maps. This is consistent with an independent inference of the CS emitting region indicating $z/r \lesssim 0.1$ using spatially-resolved T$_{\rm{rot}}$ profiles and the known 2D gas structure of the HD~163296 disk.
\end{enumerate}

The authors thank the anonymous referee for valuable comments that improved both the content and presentation of this work. This paper makes use of the following ALMA data: ADS/JAO.ALMA\#2015.1.00847.S, 2015.1.01137.S, 2016.1.00884.S, 2016.1.01086.S, 2017.1.01682.S, 2018.1.01055.L, 2021.1.00535.S, and 2021.1.00899.S. ALMA is a partnership of ESO (representing its member states), NSF (USA) and NINS (Japan), together with NRC (Canada), MOST and ASIAA (Taiwan), and KASI (Republic of Korea), in cooperation with the Republic of Chile. The Joint ALMA Observatory is operated by ESO, AUI/NRAO and NAOJ. The National Radio Astronomy Observatory is a facility of the National Science Foundation operated under cooperative agreement by Associated Universities, Inc. Support for C.J.L. was provided by NASA through the NASA Hubble Fellowship grant No. HST-HF2-51535.001-A awarded by the Space Telescope Science Institute, which is operated by the Association of Universities for Research in Astronomy, Inc., for NASA, under contract NAS5-26555. C.H.-V. acknowledges support from the National Agency for Research and Development (ANID) -- Scholarship Program through the Doctorado Nacional grant No. 2021-21212409. Y.Y. acknowledges support by Grant-in-Aid for the Japan Society for the Promotion of Science (JSPS) Fellows (KAKENHI Grant Number JP23KJ0636).

This paper also makes use of data obtained with the SMA. The Submillimeter Array is a joint project between the Smithsonian Astrophysical Observatory and the Academia Sinica Institute of Astronomy and Astrophysics and is funded by the Smithsonian Institution and the Academia Sinica. The authors wish to recognize and acknowledge the very significant cultural role and reverence that the summit of Maunakea has always had within the indigenous Hawaiian community. We are most fortunate to have had the opportunity to conduct observation from this mountain.


%

\facilities{ALMA, SMA}


\software{Astropy \citep{astropy_2013,astropy_2018}, \texttt{bettermoments} \citep{Teague18_bettermoments}, CASA \citep{McMullin_etal_2007, CASA22}, \texttt{cmasher} \citep{vanderVelden20}, \texttt{GoFish} \citep{Teague19JOSS}, \texttt{keplerian\_mask} \citep{rich_teague_2020_4321137}, Matplotlib \citep{Hunter07}, MIR (\url{https://lweb.cfa.harvard.edu/~cqi/mircook.html}), NumPy \citep{vanderWalt_etal_2011}, RADEX \citep{vanderTak07}}



\clearpage
\appendix

\section{SMA Data Calibration} \label{sec:appendix:observational details}

The HD~163296 disk was observed in the compact configuration of the SMA on 2020 Aug 10 as part of project 2020A-S018 (PI: R. Le Gal). The observations consisted of 8 antennas with a typical 225~GHz opacity of 0.16~mm and used both the 230~GHz and 240~GHz receivers of the SWARM correlator, which at the time, provided 32~GHz of bandwidth. The tuning was set to cover the CS J=4--3 (195.95421~GHz) and CS J=5--4 (244.93564~GHz) lines.

The SMA data were calibrated using the MIR software package. During calibration, all data from Antenna 8 in the 240~GHz receiver, which covered the CS J=5–4 line, was flagged due to unstable phases. The bright quasar 3c454.3 was used as the passband calibrator, and Uranus was used as the flux calibrator. Gain calibration was performed using the quasars 1743-038 and nrao530. The calibrated visibilities were then exported into CASA \texttt{v6.3} \citep{McMullin_etal_2007, CASA22} for subsequent continuum subtraction and imaging.

\section{Peak Intensity Maps} \label{sec:appendix_Fnu_maps}

\begin{figure*}[p!]
\centering
\includegraphics[width=0.97\linewidth]{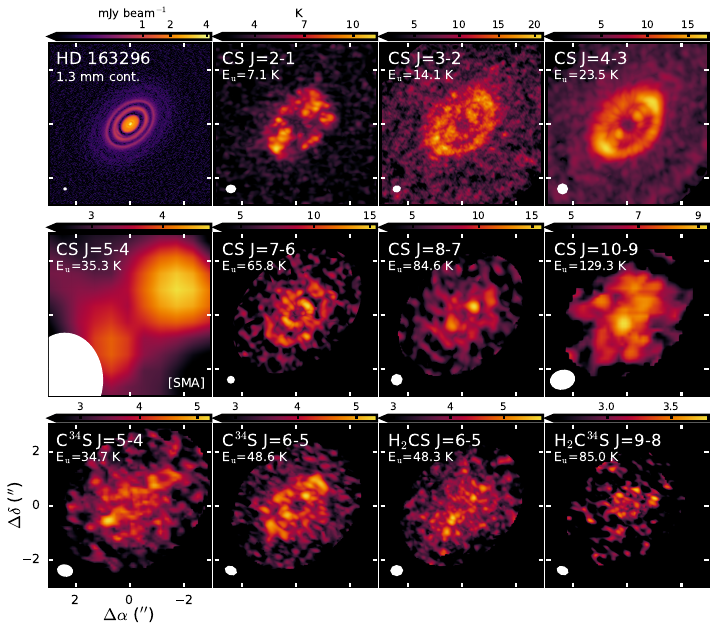}
\vspace{-8pt}
\caption{1.3~mm continuum image \citep{Andrews18, Huang18_DSHARPII} and peak intensity maps of multiple CS, C$^{34}$S, H$_2$CS, and H$_2$C$^{34}$S transitions in the HD~163296 disk. Color stretches were individually optimized and applied to each panel to increase the visibility of faint structures. The synthesized beam is shown in the lower left corner of each panel.}
\label{fig:figure_appendix_peak_intensities}
\end{figure*}

Figure \ref{fig:figure_appendix_peak_intensities} shows peak intensity maps of all lines. In a few cases, the peak intensities more clearly show the emission morphology, which is not apparent in the zeroth moment maps (Figure \ref{fig:figure1}).

\section{The Impact of Rotational Temperature on the N(H$_2$CS) / N(H$_2$C$^{34}$S) Ratio} \label{sec:appendix:H2CS_trot}

Since we only have one transition of both H$_2$CS and H$_2$C$^{34}$S, we must adopt a rotational temperature to derive column densities. In the main text, we chose to use the CS-derived T$_{\rm{rot}}$ values. Here, we explore the impact and uncertainties of this decision.

\begin{figure*}[p!]
\centering
\includegraphics[width=\linewidth]{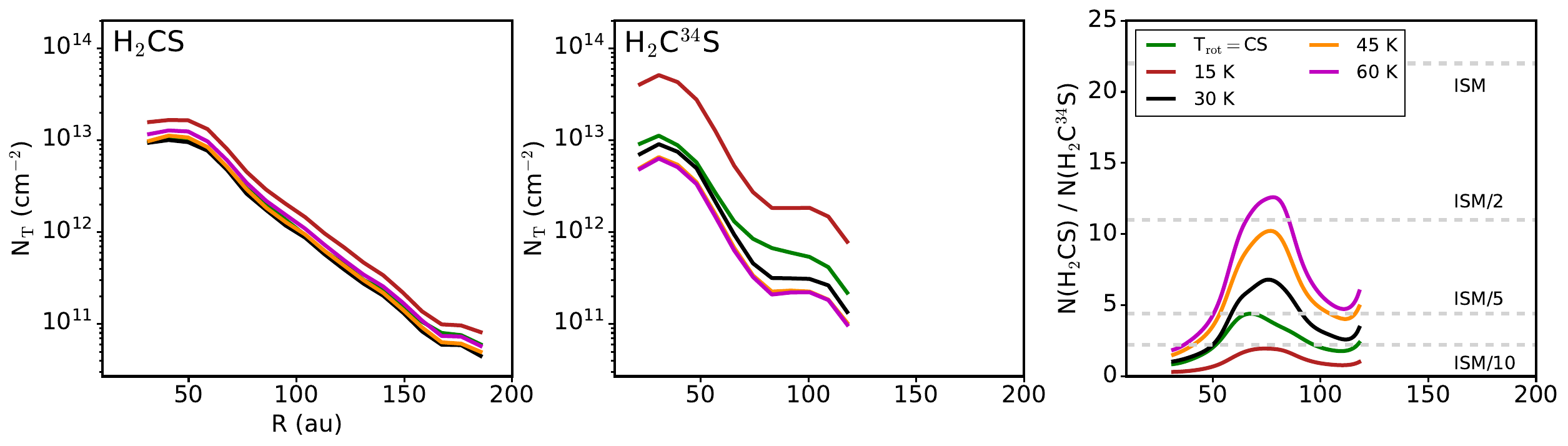}
\vspace{-18pt}
\caption{Column density profiles of H$_2$CS (\textit{left}) and H$_2$C$^{34}$S (\textit{middle}) and derived N(H$_2$CS)/N(H$_2$C$^{34}$S) ratios (\textit{right}) computed with different assumptions of T$_{\rm{rot}}$, which are shown as colored lines.}
\label{fig:figure_appendix_H2CS_Trot}
\end{figure*}

Figure \ref{fig:figure_appendix_H2CS_Trot} shows the column density profiles of H$_2$CS and H$_2$C$^{34}$S, and the correspondingly $^{32}$S/$^{34}$S ratio computed with a range of T$_{\rm{rot}}$ values. We tested a set of constant temperatures from 15~K, 30~K, 45~K, and 60~K, in addition to the CS-assumed T$_{\rm{rot}}$, which all result in a non-ISM like $^{32}$S/$^{34}$S ratio. Thus, our conclusion of significant sulfur fractionation in the H$_2$CS molecule does not depend sensitively on the assumed rotational temperature.

\clearpage


\bibliography{HD163296_CS_H2CS}{}

\begin{thebibliography}{}
\expandafter\ifx\csname natexlab\endcsname\relax\def\natexlab#1{#1}\fi
\providecommand{\url}[1]{\href{#1}{#1}}
\providecommand{\dodoi}[1]{doi:~\href{http://doi.org/#1}{\nolinkurl{#1}}}
\providecommand{\doeprint}[1]{\href{http://ascl.net/#1}{\nolinkurl{http://ascl.net/#1}}}
\providecommand{\doarXiv}[1]{\href{https://arxiv.org/abs/#1}{\nolinkurl{https://arxiv.org/abs/#1}}}

\bibitem[{{Ag{\'u}ndez} {et~al.}(2018){Ag{\'u}ndez}, {Roueff}, {Le Petit}, \& {Le Bourlot}}]{Agundez18}
{Ag{\'u}ndez}, M., {Roueff}, E., {Le Petit}, F., \& {Le Bourlot}, J. 2018, \aap, 616, A19, \dodoi{10.1051/0004-6361/201732518}

\bibitem[{{Ahrens} \& {Winnewisser}(1999)}]{Ahrens99}
{Ahrens}, V., \& {Winnewisser}, G. 1999, Zeitschrift Naturforschung Teil A, 54, 131, \dodoi{10.1515/zna-1999-0207}

\bibitem[{{Alarc{\'o}n} {et~al.}(2022){Alarc{\'o}n}, {Bergin}, \& {Teague}}]{Alarcon22}
{Alarc{\'o}n}, F., {Bergin}, E.~A., \& {Teague}, R. 2022, \apjl, 941, L24, \dodoi{10.3847/2041-8213/aca6e6}

\bibitem[{{Altwegg} {et~al.}(2022){Altwegg}, {Combi}, {Fuselier}, {H{\"a}nni}, {De Keyser}, {Mahjoub}, {M{\"u}ller}, {Pestoni}, {Rubin}, \& {Wampfler}}]{Altwegg22}
{Altwegg}, K., {Combi}, M., {Fuselier}, S.~A., {et~al.} 2022, \mnras, 516, 3900, \dodoi{10.1093/mnras/stac2440}

\bibitem[{{Andrews} {et~al.}(2018){Andrews}, {Huang}, {P{\'e}rez}, {Isella}, {Dullemond}, {Kurtovic}, {Guzm{\'a}n}, {Carpenter}, {Wilner}, {Zhang}, {Zhu}, {Birnstiel}, {Bai}, {Benisty}, {Hughes}, {{\"O}berg}, \& {Ricci}}]{Andrews18}
{Andrews}, S.~M., {Huang}, J., {P{\'e}rez}, L.~M., {et~al.} 2018, \apjl, 869, L41, \dodoi{10.3847/2041-8213/aaf741}

\bibitem[{{Artur de la Villarmois} {et~al.}(2023){Artur de la Villarmois}, {Guzm{\'a}n}, {Yang}, {Zhang}, \& {Sakai}}]{Villarmois23}
{Artur de la Villarmois}, E., {Guzm{\'a}n}, V.~V., {Yang}, Y.~L., {Zhang}, Y., \& {Sakai}, N. 2023, \aap, 678, A124, \dodoi{10.1051/0004-6361/202346728}

\bibitem[{{Astropy Collaboration} {et~al.}(2013){Astropy Collaboration}, {Robitaille}, {Tollerud}, {Greenfield}, {Droettboom}, {Bray}, {Aldcroft}, {Davis}, {Ginsburg}, {Price-Whelan}, {Kerzendorf}, {Conley}, {Crighton}, {Barbary}, {Muna}, {Ferguson}, {Grollier}, {Parikh}, {Nair}, {Unther}, {Deil}, {Woillez}, {Conseil}, {Kramer}, {Turner}, {Singer}, {Fox}, {Weaver}, {Zabalza}, {Edwards}, {Azalee Bostroem}, {Burke}, {Casey}, {Crawford}, {Dencheva}, {Ely}, {Jenness}, {Labrie}, {Lim}, {Pierfederici}, {Pontzen}, {Ptak}, {Refsdal}, {Servillat}, \& {Streicher}}]{astropy_2013}
{Astropy Collaboration}, {Robitaille}, T.~P., {Tollerud}, E.~J., {et~al.} 2013, \aap, 558, A33, \dodoi{10.1051/0004-6361/201322068}

\bibitem[{{Bergin} {et~al.}(2024){Bergin}, {Bosman}, {Teague}, {Calahan}, {Willacy}, {Cleeves}, {Schwarz}, {Zhang}, \& {Bruderer}}]{Bergin24}
{Bergin}, E.~A., {Bosman}, A., {Teague}, R., {et~al.} 2024, \apj, 965, 147, \dodoi{10.3847/1538-4357/ad3443}

\bibitem[{{Bergner} {et~al.}(2019){Bergner}, {{\"O}berg}, {Bergin}, {Loomis}, {Pegues}, \& {Qi}}]{Bergner19}
{Bergner}, J.~B., {{\"O}berg}, K.~I., {Bergin}, E.~A., {et~al.} 2019, \apj, 876, 25, \dodoi{10.3847/1538-4357/ab141e}

\bibitem[{{Bergner} {et~al.}(2021){Bergner}, {{\"O}berg}, {Guzm{\'a}n}, {Law}, {Loomis}, {Cataldi}, {Bosman}, {Aikawa}, {Andrews}, {Bergin}, {Booth}, {Cleeves}, {Czekala}, {Huang}, {Ilee}, {Le Gal}, {Long}, {Nomura}, {M{\'e}nard}, {Qi}, {Schwarz}, {Teague}, {Tsukagoshi}, {Walsh}, {Wilner}, \& {Yamato}}]{Bergner21ApJS}
{Bergner}, J.~B., {{\"O}berg}, K.~I., {Guzm{\'a}n}, V.~V., {et~al.} 2021, \apjs, 257, 11, \dodoi{10.3847/1538-4365/ac143a}

\bibitem[{{Biver} {et~al.}(2016){Biver}, {Moreno}, {Bockel{\'e}e-Morvan}, {Sandqvist}, {Colom}, {Crovisier}, {Lis}, {Boissier}, {Debout}, {Paubert}, {Milam}, {Hjalmarson}, {Lundin}, {Karlsson}, {Battelino}, {Frisk}, {Murtagh}, \& {Odin Team}}]{Biver16}
{Biver}, N., {Moreno}, R., {Bockel{\'e}e-Morvan}, D., {et~al.} 2016, \aap, 589, A78, \dodoi{10.1051/0004-6361/201528041}

\bibitem[{{Bockel{\'e}e-Morvan} {et~al.}(2004){Bockel{\'e}e-Morvan}, {Crovisier}, {Mumma}, \& {Weaver}}]{BM04}
{Bockel{\'e}e-Morvan}, D., {Crovisier}, J., {Mumma}, M.~J., \& {Weaver}, H.~A. 2004, in Comets II, ed. M.~C. {Festou}, H.~U. {Keller}, \& H.~A. {Weaver}, 391

\bibitem[{{Bogey} {et~al.}(1981){Bogey}, {Demuynck}, \& {Destombes}}]{Bogey81}
{Bogey}, M., {Demuynck}, C., \& {Destombes}, J.~L. 1981, Chemical Physics Letters, 81, 256, \dodoi{10.1016/0009-2614(81)80247-3}

\bibitem[{{Bogey} {et~al.}(1982){Bogey}, {Demuynck}, \& {Destombes}}]{Bogey82}
---. 1982, Journal of Molecular Spectroscopy, 95, 35, \dodoi{10.1016/0022-2852(82)90234-X}

\bibitem[{{Boogert} {et~al.}(2015){Boogert}, {Gerakines}, \& {Whittet}}]{Boogert15}
{Boogert}, A.~C.~A., {Gerakines}, P.~A., \& {Whittet}, D. C.~B. 2015, \araa, 53, 541, \dodoi{10.1146/annurev-astro-082214-122348}

\bibitem[{{Booth} {et~al.}(2023{\natexlab{a}}){Booth}, {Ilee}, {Walsh}, {Kama}, {Keyte}, {van Dishoeck}, \& {Nomura}}]{Booth23}
{Booth}, A.~S., {Ilee}, J.~D., {Walsh}, C., {et~al.} 2023{\natexlab{a}}, \aap, 669, A53, \dodoi{10.1051/0004-6361/202244472}

\bibitem[{{Booth} {et~al.}(2023{\natexlab{b}}){Booth}, {Law}, {Temmink}, {Leemker}, \& {Mac{\'\i}as}}]{Booth23_HD169142}
{Booth}, A.~S., {Law}, C.~J., {Temmink}, M., {Leemker}, M., \& {Mac{\'\i}as}, E. 2023{\natexlab{b}}, \aap, 678, A146, \dodoi{10.1051/0004-6361/202346974}

\bibitem[{{Booth} {et~al.}(2021){Booth}, {Tabone}, {Ilee}, {Walsh}, {Aikawa}, {Andrews}, {Bae}, {Bergin}, {Bergner}, {Bosman}, {Calahan}, {Cataldi}, {Cleeves}, {Czekala}, {Guzm{\'a}n}, {Huang}, {Law}, {Le Gal}, {Long}, {Loomis}, {M{\'e}nard}, {Nomura}, {{\"O}berg}, {Qi}, {Schwarz}, {Teague}, {Tsukagoshi}, {Wilner}, {Yamato}, \& {Zhang}}]{Booth21_MAPS}
{Booth}, A.~S., {Tabone}, B., {Ilee}, J.~D., {et~al.} 2021, \apjs, 257, 16, \dodoi{10.3847/1538-4365/ac1ad4}

\bibitem[{{Booth} {et~al.}(2024{\natexlab{a}}){Booth}, {Leemker}, {van Dishoeck}, {Evans}, {Ilee}, {Kama}, {Keyte}, {Law}, {van der Marel}, {Nomura}, {Notsu}, {{\"O}berg}, {Temmink}, \& {Walsh}}]{Booth24_HD100546}
{Booth}, A.~S., {Leemker}, M., {van Dishoeck}, E.~F., {et~al.} 2024{\natexlab{a}}, \aj, 167, 164, \dodoi{10.3847/1538-3881/ad2700}

\bibitem[{{Booth} {et~al.}(2024{\natexlab{b}}){Booth}, {Temmink}, {van Dishoeck}, {Evans}, {Ilee}, {Kama}, {Keyte}, {Law}, {Leemker}, {van der Marel}, {Nomura}, {Notsu}, {{\"O}berg}, \& {Walsh}}]{Booth24_IRS48}
{Booth}, A.~S., {Temmink}, M., {van Dishoeck}, E.~F., {et~al.} 2024{\natexlab{b}}, \aj, 167, 165, \dodoi{10.3847/1538-3881/ad26ff}

\bibitem[{{Booth} {et~al.}(2024{\natexlab{c}}){Booth}, {Drozdovskaya}, {Temmink}, {Nomura}, {van Dishoeck}, {Keyte}, {Law}, {Leemker}, {van der Marel}, {Notsu}, {{\"O}berg}, \& {Walsh}}]{Booth24}
{Booth}, A.~S., {Drozdovskaya}, M.~N., {Temmink}, M., {et~al.} 2024{\natexlab{c}}, \apj, 975, 72, \dodoi{10.3847/1538-4357/ad7817}

\bibitem[{{Bosman} \& {Bergin}(2021)}]{Bosman21}
{Bosman}, A.~D., \& {Bergin}, E.~A. 2021, \apjl, 918, L10, \dodoi{10.3847/2041-8213/ac1db1}

\bibitem[{{Calahan} {et~al.}(2021){Calahan}, {Bergin}, {Zhang}, {Schwarz}, {{\"O}berg}, {Guzm{\'a}n}, {Walsh}, {Aikawa}, {Alarc{\'o}n}, {Andrews}, {Bae}, {Bergner}, {Booth}, {Bosman}, {Cataldi}, {Czekala}, {Huang}, {Ilee}, {Law}, {Le Gal}, {Long}, {Loomis}, {M{\'e}nard}, {Nomura}, {Qi}, {Teague}, {van't Hoff}, {Wilner}, \& {Yamato}}]{Calahan21}
{Calahan}, J.~K., {Bergin}, E.~A., {Zhang}, K., {et~al.} 2021, \apjs, 257, 17, \dodoi{10.3847/1538-4365/ac143f}

\bibitem[{{Calcino} {et~al.}(2022){Calcino}, {Hilder}, {Price}, {Pinte}, {Bollati}, {Lodato}, \& {Norfolk}}]{Calcino22}
{Calcino}, J., {Hilder}, T., {Price}, D.~J., {et~al.} 2022, \apjl, 929, L25, \dodoi{10.3847/2041-8213/ac64a7}

\bibitem[{{Calmonte} {et~al.}(2016){Calmonte}, {Altwegg}, {Balsiger}, {Berthelier}, {Bieler}, {Cessateur}, {Dhooghe}, {van Dishoeck}, {Fiethe}, {Fuselier}, {Gasc}, {Gombosi}, {H{\"a}ssig}, {Le Roy}, {Rubin}, {S{\'e}mon}, {Tzou}, \& {Wampfler}}]{Calmonte16}
{Calmonte}, U., {Altwegg}, K., {Balsiger}, H., {et~al.} 2016, \mnras, 462, S253, \dodoi{10.1093/mnras/stw2601}

\bibitem[{{Calmonte} {et~al.}(2017){Calmonte}, {Altwegg}, {Balsiger}, {Berthelier}, {Bieler}, {De Keyser}, {Fiethe}, {Fuselier}, {Gasc}, {Gombosi}, {Le Roy}, {Rubin}, {S{\'e}mon}, {Tzou}, \& {Wampfler}}]{Calmonte17}
---. 2017, \mnras, 469, S787, \dodoi{10.1093/mnras/stx2534}

\bibitem[{{CASA Team} {et~al.}(2022){CASA Team}, {Bean}, {Bhatnagar}, {Castro}, {Donovan Meyer}, {Emonts}, {Garcia}, {Garwood}, {Golap}, {Gonzalez Villalba}, {Harris}, {Hayashi}, {Hoskins}, {Hsieh}, {Jagannathan}, {Kawasaki}, {Keimpema}, {Kettenis}, {Lopez}, {Marvil}, {Masters}, {McNichols}, {Mehringer}, {Miel}, {Moellenbrock}, {Montesino}, {Nakazato}, {Ott}, {Petry}, {Pokorny}, {Raba}, {Rau}, {Schiebel}, {Schweighart}, {Sekhar}, {Shimada}, {Small}, {Steeb}, {Sugimoto}, {Suoranta}, {Tsutsumi}, {van Bemmel}, {Verkouter}, {Wells}, {Xiong}, {Szomoru}, {Griffith}, {Glendenning}, \& {Kern}}]{CASA22}
{CASA Team}, {Bean}, B., {Bhatnagar}, S., {et~al.} 2022, \pasp, 134, 114501, \dodoi{10.1088/1538-3873/ac9642}

\bibitem[{{Cataldi} {et~al.}(2021){Cataldi}, {Yamato}, {Aikawa}, {Bergner}, {Furuya}, {Guzm{\'a}n}, {Huang}, {Loomis}, {Qi}, {Andrews}, {Bergin}, {Booth}, {Bosman}, {Cleeves}, {Czekala}, {Ilee}, {Law}, {Le Gal}, {Liu}, {Long}, {M{\'e}nard}, {Nomura}, {{\"O}berg}, {Schwarz}, {Teague}, {Tsukagoshi}, {Walsh}, {Wilner}, \& {Zhang}}]{Cataldi21}
{Cataldi}, G., {Yamato}, Y., {Aikawa}, Y., {et~al.} 2021, \apjs, 257, 10, \dodoi{10.3847/1538-4365/ac143d}

\bibitem[{{Chen} {et~al.}(2015){Chen}, {Juang}, {Nuevo}, {Jim{\'e}nez-Escobar}, {Mu{\~n}oz Caro}, {Qiu}, {Chu}, {Yih}, {Wu}, {Fung}, \& {Ip}}]{Chen15}
{Chen}, Y.~J., {Juang}, K.~J., {Nuevo}, M., {et~al.} 2015, \apj, 798, 80, \dodoi{10.1088/0004-637X/798/2/80}

\bibitem[{{Czekala} {et~al.}(2021){Czekala}, {Loomis}, {Teague}, {Booth}, {Huang}, {Cataldi}, {Ilee}, {Law}, {Walsh}, {Bosman}, {Guzm{\'a}n}, {Gal}, {{\"O}berg}, {Yamato}, {Aikawa}, {Andrews}, {Bae}, {Bergin}, {Bergner}, {Cleeves}, {Kurtovic}, {M{\'e}nard}, {Nomura}, {P{\'e}rez}, {Qi}, {Schwarz}, {Tsukagoshi}, {Waggoner}, {Wilner}, \& {Zhang}}]{Czekala21}
{Czekala}, I., {Loomis}, R.~A., {Teague}, R., {et~al.} 2021, \apjs, 257, 2, \dodoi{10.3847/1538-4365/ac1430}

\bibitem[{{Doi} \& {Kataoka}(2021)}]{Doi21}
{Doi}, K., \& {Kataoka}, A. 2021, \apj, 912, 164, \dodoi{10.3847/1538-4357/abe5a6}

\bibitem[{{Dutrey} {et~al.}(1997){Dutrey}, {Guilloteau}, \& {Guelin}}]{Dutrey97}
{Dutrey}, A., {Guilloteau}, S., \& {Guelin}, M. 1997, \aap, 317, L55

\bibitem[{{Dutrey} {et~al.}(2011){Dutrey}, {Wakelam}, {Boehler}, {Guilloteau}, {Hersant}, {Semenov}, {Chapillon}, {Henning}, {Pi{\'e}tu}, {Launhardt}, {Gueth}, \& {Schreyer}}]{Dutrey11}
{Dutrey}, A., {Wakelam}, V., {Boehler}, Y., {et~al.} 2011, \aap, 535, A104, \dodoi{10.1051/0004-6361/201116931}

\bibitem[{{Dutrey} {et~al.}(2024){Dutrey}, {Chapillon}, {Guilloteau}, {Tang}, {Boccaletti}, {Bouscasse}, {Collin-Dufresne}, {Di Folco}, {Fuente}, {Pi{\'e}tu}, {Rivi{\`e}re-Marichalar}, \& {Semenov}}]{Dutrey24}
{Dutrey}, A., {Chapillon}, E., {Guilloteau}, S., {et~al.} 2024, \aap, 689, L7, \dodoi{10.1051/0004-6361/202451299}

\bibitem[{{Endres} {et~al.}(2016){Endres}, {Schlemmer}, {Schilke}, {Stutzki}, \& {M{\"u}ller}}]{Endres16}
{Endres}, C.~P., {Schlemmer}, S., {Schilke}, P., {Stutzki}, J., \& {M{\"u}ller}, H. S.~P. 2016, Journal of Molecular Spectroscopy, 327, 95, \dodoi{10.1016/j.jms.2016.03.005}

\bibitem[{{Fairlamb} {et~al.}(2015){Fairlamb}, {Oudmaijer}, {Mendigut{\'\i}a}, {Ilee}, \& {van den Ancker}}]{Fairlamb15}
{Fairlamb}, J.~R., {Oudmaijer}, R.~D., {Mendigut{\'\i}a}, I., {Ilee}, J.~D., \& {van den Ancker}, M.~E. 2015, \mnras, 453, 976, \dodoi{10.1093/mnras/stv1576}

\bibitem[{{Favre} {et~al.}(2015){Favre}, {Bergin}, {Cleeves}, {Hersant}, {Qi}, \& {Aikawa}}]{Favre15}
{Favre}, C., {Bergin}, E.~A., {Cleeves}, L.~I., {et~al.} 2015, \apjl, 802, L23, \dodoi{10.1088/2041-8205/802/2/L23}

\bibitem[{{Fedele} {et~al.}(2012){Fedele}, {Bruderer}, {van Dishoeck}, {Herczeg}, {Evans}, {Bouwman}, {Henning}, \& {Green}}]{Fedele12}
{Fedele}, D., {Bruderer}, S., {van Dishoeck}, E.~F., {et~al.} 2012, \aap, 544, L9, \dodoi{10.1051/0004-6361/201219615}

\bibitem[{{Flaherty} {et~al.}(2017){Flaherty}, {Hughes}, {Rose}, {Simon}, {Qi}, {Andrews}, {K{\'o}sp{\'a}l}, {Wilner}, {Chiang}, {Armitage}, \& {Bai}}]{Flaherty17}
{Flaherty}, K.~M., {Hughes}, A.~M., {Rose}, S.~C., {et~al.} 2017, \apj, 843, 150, \dodoi{10.3847/1538-4357/aa79f9}

\bibitem[{{Foreman-Mackey} {et~al.}(2013){Foreman-Mackey}, {Hogg}, {Lang}, \& {Goodman}}]{Foreman_Mackey13}
{Foreman-Mackey}, D., {Hogg}, D.~W., {Lang}, D., \& {Goodman}, J. 2013, \pasp, 125, 306, \dodoi{10.1086/670067}

\bibitem[{{Fuente} {et~al.}(2010){Fuente}, {Cernicharo}, {Ag{\'u}ndez}, {Bern{\'e}}, {Goicoechea}, {Alonso-Albi}, \& {Marcelino}}]{Fuente10}
{Fuente}, A., {Cernicharo}, J., {Ag{\'u}ndez}, M., {et~al.} 2010, \aap, 524, A19, \dodoi{10.1051/0004-6361/201014905}

\bibitem[{{Fuente} {et~al.}(2023){Fuente}, {Rivi{\`e}re-Marichalar}, {Beitia-Antero}, {Caselli}, {Wakelam}, {Esplugues}, {Rodr{\'\i}guez-Baras}, {Navarro-Almaida}, {Gerin}, {Kramer}, {Bachiller}, {Goicoechea}, {Jim{\'e}nez-Serra}, {Loison}, {Ivlev}, {Mart{\'\i}n-Dom{\'e}nech}, {Spezzano}, {Roncero}, {Mu{\~n}oz-Caro}, {Cazaux}, \& {Marcelino}}]{Fuente23}
{Fuente}, A., {Rivi{\`e}re-Marichalar}, P., {Beitia-Antero}, L., {et~al.} 2023, \aap, 670, A114, \dodoi{10.1051/0004-6361/202244843}

\bibitem[{{Furuya} {et~al.}(2022){Furuya}, {Tsukagoshi}, {Qi}, {Nomura}, {Cleeves}, {Lee}, \& {Yoshida}}]{Furuya22}
{Furuya}, K., {Tsukagoshi}, T., {Qi}, C., {et~al.} 2022, \apj, 926, 148, \dodoi{10.3847/1538-4357/ac45ff}

\bibitem[{{Gaia Collaboration} {et~al.}(2021){Gaia Collaboration}, {Brown}, {Vallenari}, {Prusti}, {de Bruijne}, {Babusiaux}, {Biermann}, {Creevey}, {Evans}, {Eyer}, {Hutton}, {Jansen}, {Jordi}, {Klioner}, {Lammers}, {Lindegren}, {Luri}, {Mignard}, {Panem}, {Pourbaix}, {Randich}, {Sartoretti}, {Soubiran}, {Walton}, {Arenou}, {Bailer-Jones}, {Bastian}, {Cropper}, {Drimmel}, {Katz}, {Lattanzi}, {van Leeuwen}, {Bakker}, {Cacciari}, {Casta{\~n}eda}, {De Angeli}, {Ducourant}, {Fabricius}, {Fouesneau}, {Fr{\'e}mat}, {Guerra}, {Guerrier}, {Guiraud}, {Jean-Antoine Piccolo}, {Masana}, {Messineo}, {Mowlavi}, {Nicolas}, {Nienartowicz}, {Pailler}, {Panuzzo}, {Riclet}, {Roux}, {Seabroke}, {Sordo}, {Tanga}, {Th{\'e}venin}, {Gracia-Abril}, {Portell}, {Teyssier}, {Altmann}, {Andrae}, {Bellas-Velidis}, {Benson}, {Berthier}, {Blomme}, {Brugaletta}, {Burgess}, {Busso}, {Carry}, {Cellino}, {Cheek}, {Clementini}, {Damerdji}, {Davidson}, {Delchambre}, {Dell'Oro}, {Fern{\'a}ndez-Hern{\'a}ndez}, {Galluccio}, {Garc{\'\i}a-Lario},
  {Garcia-Reinaldos}, {Gonz{\'a}lez-N{\'u}{\~n}ez}, {Gosset}, {Haigron}, {Halbwachs}, {Hambly}, {Harrison}, {Hatzidimitriou}, {Heiter}, {Hern{\'a}ndez}, {Hestroffer}, {Hodgkin}, {Holl}, {Jan{\ss}en}, {Jevardat de Fombelle}, {Jordan}, {Krone-Martins}, {Lanzafame}, {L{\"o}ffler}, {Lorca}, {Manteiga}, {Marchal}, {Marrese}, {Moitinho}, {Mora}, {Muinonen}, {Osborne}, {Pancino}, {Pauwels}, {Petit}, {Recio-Blanco}, {Richards}, {Riello}, {Rimoldini}, {Robin}, {Roegiers}, {Rybizki}, {Sarro}, {Siopis}, {Smith}, {Sozzetti}, {Ulla}, {Utrilla}, {van Leeuwen}, {van Reeven}, {Abbas}, {Abreu Aramburu}, {Accart}, {Aerts}, {Aguado}, {Ajaj}, {Altavilla}, {{\'A}lvarez}, {{\'A}lvarez Cid-Fuentes}, {Alves}, {Anderson}, {Anglada Varela}, {Antoja}, {Audard}, {Baines}, {Baker}, {Balaguer-N{\'u}{\~n}ez}, {Balbinot}, {Balog}, {Barache}, {Barbato}, {Barros}, {Barstow}, {Bartolom{\'e}}, {Bassilana}, {Bauchet}, {Baudesson-Stella}, {Becciani}, {Bellazzini}, {Bernet}, {Bertone}, {Bianchi}, {Blanco-Cuaresma}, {Boch}, {Bombrun}, {Bossini},
  {Bouquillon}, {Bragaglia}, {Bramante}, {Breedt}, {Bressan}, {Brouillet}, {Bucciarelli}, {Burlacu}, {Busonero}, {Butkevich}, {Buzzi}, {Caffau}, {Cancelliere}, {C{\'a}novas}, {Cantat-Gaudin}, {Carballo}, {Carlucci}, {Carnerero}, {Carrasco}, {Casamiquela}, {Castellani}, {Castro-Ginard}, {Castro Sampol}, {Chaoul}, {Charlot}, {Chemin}, {Chiavassa}, {Cioni}, {Comoretto}, {Cooper}, {Cornez}, {Cowell}, {Crifo}, {Crosta}, {Crowley}, {Dafonte}, {Dapergolas}, {David}, {David}, {de Laverny}, {De Luise}, {De March}, {De Ridder}, {de Souza}, {de Teodoro}, {de Torres}, {del Peloso}, {del Pozo}, {Delbo}, {Delgado}, {Delgado}, {Delisle}, {Di Matteo}, {Diakite}, {Diener}, {Distefano}, {Dolding}, {Eappachen}, {Edvardsson}, {Enke}, {Esquej}, {Fabre}, {Fabrizio}, {Faigler}, {Fedorets}, {Fernique}, {Fienga}, {Figueras}, {Fouron}, {Fragkoudi}, {Fraile}, {Franke}, {Gai}, {Garabato}, {Garcia-Gutierrez}, {Garc{\'\i}a-Torres}, {Garofalo}, {Gavras}, {Gerlach}, {Geyer}, {Giacobbe}, {Gilmore}, {Girona}, {Giuffrida}, {Gomel}, {Gomez},
  {Gonzalez-Santamaria}, {Gonz{\'a}lez-Vidal}, {Granvik}, {Guti{\'e}rrez-S{\'a}nchez}, {Guy}, {Hauser}, {Haywood}, {Helmi}, {Hidalgo}, {Hilger}, {H{\l}adczuk}, {Hobbs}, {Holland}, {Huckle}, {Jasniewicz}, {Jonker}, {Juaristi Campillo}, {Julbe}, {Karbevska}, {Kervella}, {Khanna}, {Kochoska}, {Kontizas}, {Kordopatis}, {Korn}, {Kostrzewa-Rutkowska}, {Kruszy{\'n}ska}, {Lambert}, {Lanza}, {Lasne}, {Le Campion}, {Le Fustec}, {Lebreton}, {Lebzelter}, {Leccia}, {Leclerc}, {Lecoeur-Taibi}, {Liao}, {Licata}, {Lindstr{\o}m}, {Lister}, {Livanou}, {Lobel}, {Madrero Pardo}, {Managau}, {Mann}, {Marchant}, {Marconi}, {Marcos Santos}, {Marinoni}, {Marocco}, {Marshall}, {Martin Polo}, {Mart{\'\i}n-Fleitas}, {Masip}, {Massari}, {Mastrobuono-Battisti}, {Mazeh}, {McMillan}, {Messina}, {Michalik}, {Millar}, {Mints}, {Molina}, {Molinaro}, {Moln{\'a}r}, {Montegriffo}, {Mor}, {Morbidelli}, {Morel}, {Morris}, {Mulone}, {Munoz}, {Muraveva}, {Murphy}, {Musella}, {Noval}, {Ord{\'e}novic}, {Orr{\`u}}, {Osinde}, {Pagani}, {Pagano},
  {Palaversa}, {Palicio}, {Panahi}, {Pawlak}, {Pe{\~n}alosa Esteller}, {Penttil{\"a}}, {Piersimoni}, {Pineau}, {Plachy}, {Plum}, {Poggio}, {Poretti}, {Poujoulet}, {Pr{\v{s}}a}, {Pulone}, {Racero}, {Ragaini}, {Rainer}, {Raiteri}, {Rambaux}, {Ramos}, {Ramos-Lerate}, {Re Fiorentin}, {Regibo}, {Reyl{\'e}}, {Ripepi}, {Riva}, {Rixon}, {Robichon}, {Robin}, {Roelens}, {Rohrbasser}, {Romero-G{\'o}mez}, {Rowell}, {Royer}, {Rybicki}, {Sadowski}, {Sagrist{\`a} Sell{\'e}s}, {Sahlmann}, {Salgado}, {Salguero}, {Samaras}, {Sanchez Gimenez}, {Sanna}, {Santove{\~n}a}, {Sarasso}, {Schultheis}, {Sciacca}, {Segol}, {Segovia}, {S{\'e}gransan}, {Semeux}, {Shahaf}, {Siddiqui}, {Siebert}, {Siltala}, {Slezak}, {Smart}, {Solano}, {Solitro}, {Souami}, {Souchay}, {Spagna}, {Spoto}, {Steele}, {Steidelm{\"u}ller}, {Stephenson}, {S{\"u}veges}, {Szabados}, {Szegedi-Elek}, {Taris}, {Tauran}, {Taylor}, {Teixeira}, {Thuillot}, {Tonello}, {Torra}, {Torra}, {Turon}, {Unger}, {Vaillant}, {van Dillen}, {Vanel}, {Vecchiato}, {Viala}, {Vicente},
  {Voutsinas}, {Weiler}, {Wevers}, {Wyrzykowski}, {Yoldas}, {Yvard}, {Zhao}, {Zorec}, {Zucker}, {Zurbach}, \& {Zwitter}}]{Gaia21}
{Gaia Collaboration}, {Brown}, A.~G.~A., {Vallenari}, A., {et~al.} 2021, \aap, 649, A1, \dodoi{10.1051/0004-6361/202039657}

\bibitem[{{Garufi} {et~al.}(2022){Garufi}, {Podio}, {Codella}, {Segura-Cox}, {Vander Donckt}, {Mercimek}, {Bacciotti}, {Fedele}, {Kasper}, {Pineda}, {Humphreys}, \& {Testi}}]{Garufi22}
{Garufi}, A., {Podio}, L., {Codella}, C., {et~al.} 2022, \aap, 658, A104, \dodoi{10.1051/0004-6361/202141264}

\bibitem[{{Goldsmith} \& {Langer}(1999)}]{Goldsmith99}
{Goldsmith}, P.~F., \& {Langer}, W.~D. 1999, \apj, 517, 209, \dodoi{10.1086/307195}

\bibitem[{{Gottlieb} {et~al.}(2003){Gottlieb}, {Myers}, \& {Thaddeus}}]{Gottlieb03}
{Gottlieb}, C.~A., {Myers}, P.~C., \& {Thaddeus}, P. 2003, \apj, 588, 655, \dodoi{10.1086/368378}

\bibitem[{{Grady} {et~al.}(2000){Grady}, {Devine}, {Woodgate}, {Kimble}, {Bruhweiler}, {Boggess}, {Linsky}, {Plait}, {Clampin}, \& {Kalas}}]{Grady00}
{Grady}, C.~A., {Devine}, D., {Woodgate}, B., {et~al.} 2000, \apj, 544, 895, \dodoi{10.1086/317222}

\bibitem[{{Gravity Collaboration} {et~al.}(2023){Gravity Collaboration}, {Soulain}, {Perraut}, {Bouvier}, {Pantolmos}, {Caratti O Garatti}, {Caselli}, {Garcia}, {Lopez}, {Aimar}, {Amorin}, {Benisty}, {Berger}, {Bourdarot}, {Brandner}, {Cl{\'e}net}, {de Zeeuw}, {Davies}, {Drescher}, {Eckart}, {Eisenhauer}, {Schreiber}, {Gendron}, {Genzuel}, {Gillessen}, {Hei{\ss}el}, {Henning}, {Hippler}, {Horrobin}, {Jocou}, {Kervella}, {Labadie}, {Lacour}, {Lapeyrere}, {Le Bouquin}, {L{\'e}na}, {Lutz}, {Mang}, {Ott}, {Paumard}, {Perrin}, {Sanchez}, {Scheithauer}, {Shangguan}, {Shimizu}, {Straub}, {Straubmeier}, {Sturm}, {Tacconi}, {Vincent}, {van Dishoeck}, {Widmann}, {Wieprecht}, {Wiezorrek}, \& {Yazici}}]{Gravity_CITau}
{Gravity Collaboration}, {Soulain}, A., {Perraut}, K., {et~al.} 2023, \aap, 674, A203, \dodoi{10.1051/0004-6361/202346446}

\bibitem[{{Guidi} {et~al.}(2022){Guidi}, {Isella}, {Testi}, {Chandler}, {Liu}, {Schmid}, {Rosotti}, {Meng}, {Jennings}, {Williams}, {Carpenter}, {de Gregorio-Monsalvo}, {Li}, {Liu}, {Ortolani}, {Quanz}, {Ricci}, \& {Tazzari}}]{Guidi22}
{Guidi}, G., {Isella}, A., {Testi}, L., {et~al.} 2022, \aap, 664, A137, \dodoi{10.1051/0004-6361/202142303}

\bibitem[{{Guilloteau} {et~al.}(2016){Guilloteau}, {Reboussin}, {Dutrey}, {Chapillon}, {Wakelam}, {Pi{\'e}tu}, {Di Folco}, {Semenov}, \& {Henning}}]{Guilloteau16}
{Guilloteau}, S., {Reboussin}, L., {Dutrey}, A., {et~al.} 2016, \aap, 592, A124, \dodoi{10.1051/0004-6361/201527088}

\bibitem[{{Guzm{\'a}n} {et~al.}(2021){Guzm{\'a}n}, {Bergner}, {Law}, {{\"O}berg}, {Walsh}, {Cataldi}, {Aikawa}, {Bergin}, {Czekala}, {Huang}, {Andrews}, {Loomis}, {Zhang}, {Le Gal}, {Alarc{\'o}n}, {Ilee}, {Teague}, {Cleeves}, {Wilner}, {Long}, {Schwarz}, {Bosman}, {P{\'e}rez}, {M{\'e}nard}, \& {Liu}}]{Guzman21_MAPS}
{Guzm{\'a}n}, V.~V., {Bergner}, J.~B., {Law}, C.~J., {et~al.} 2021, \apjs, 257, 6, \dodoi{10.3847/1538-4365/ac1440}

\bibitem[{{Heck} {et~al.}(2012){Heck}, {Hoppe}, \& {Huth}}]{Heck12}
{Heck}, P.~R., {Hoppe}, P., \& {Huth}, J. 2012, \maps, 47, 649, \dodoi{10.1111/j.1945-5100.2012.01362.x}

\bibitem[{{Hern{\'a}ndez-Vera} {et~al.}(2024){Hern{\'a}ndez-Vera}, {Guzm{\'a}n}, {Artur de la Villarmois}, {{\"O}berg}, {Cleeves}, {Hogerheijde}, {Qi}, {Carpenter}, \& {Fayolle}}]{Hernandez_Vera24}
{Hern{\'a}ndez-Vera}, C., {Guzm{\'a}n}, V.~V., {Artur de la Villarmois}, E., {et~al.} 2024, \apj, 967, 68, \dodoi{10.3847/1538-4357/ad3cdb}

\bibitem[{{Huang} {et~al.}(2024){Huang}, {Bergin}, {Le Gal}, {Andrews}, {Bae}, {Keyte}, \& {Sturm}}]{Huang24}
{Huang}, J., {Bergin}, E.~A., {Le Gal}, R., {et~al.} 2024, \apj, 973, 135, \dodoi{10.3847/1538-4357/ad6447}

\bibitem[{{Huang} {et~al.}(2018){Huang}, {Andrews}, {Dullemond}, {Isella}, {P{\'e}rez}, {Guzm{\'a}n}, {{\"O}berg}, {Zhu}, {Zhang}, {Bai}, {Benisty}, {Birnstiel}, {Carpenter}, {Hughes}, {Ricci}, {Weaver}, \& {Wilner}}]{Huang18_DSHARPII}
{Huang}, J., {Andrews}, S.~M., {Dullemond}, C.~P., {et~al.} 2018, \apjl, 869, L42, \dodoi{10.3847/2041-8213/aaf740}

\bibitem[{{Hunter}(2007)}]{Hunter07}
{Hunter}, J.~D. 2007, Computing in Science and Engineering, 9, 90, \dodoi{10.1109/MCSE.2007.55}

\bibitem[{{Ilee} {et~al.}(2021){Ilee}, {Walsh}, {Booth}, {Aikawa}, {Andrews}, {Bae}, {Bergin}, {Bergner}, {Bosman}, {Cataldi}, {Cleeves}, {Czekala}, {Guzm{\'a}n}, {Huang}, {Law}, {Le Gal}, {Loomis}, {M{\'e}nard}, {Nomura}, {{\"O}berg}, {Qi}, {Schwarz}, {Teague}, {Tsukagoshi}, {Wilner}, {Yamato}, \& {Zhang}}]{Ilee21}
{Ilee}, J.~D., {Walsh}, C., {Booth}, A.~S., {et~al.} 2021, \apjs, 257, 9, \dodoi{10.3847/1538-4365/ac1441}

\bibitem[{{Isella} {et~al.}(2016){Isella}, {Guidi}, {Testi}, {Liu}, {Li}, {Li}, {Weaver}, {Boehler}, {Carperter}, {De Gregorio-Monsalvo}, {Manara}, {Natta}, {P{\'e}rez}, {Ricci}, {Sargent}, {Tazzari}, \& {Turner}}]{Isella16}
{Isella}, A., {Guidi}, G., {Testi}, L., {et~al.} 2016, \prl, 117, 251101, \dodoi{10.1103/PhysRevLett.117.251101}

\bibitem[{{Izquierdo} {et~al.}(2022){Izquierdo}, {Facchini}, {Rosotti}, {van Dishoeck}, \& {Testi}}]{Izquierdo22}
{Izquierdo}, A.~F., {Facchini}, S., {Rosotti}, G.~P., {van Dishoeck}, E.~F., \& {Testi}, L. 2022, \apj, 928, 2, \dodoi{10.3847/1538-4357/ac474d}

\bibitem[{{Jewitt} {et~al.}(1997){Jewitt}, {Matthews}, {Owen}, \& {Meier}}]{Jewitt97}
{Jewitt}, D., {Matthews}, H.~E., {Owen}, T., \& {Meier}, R. 1997, Science, 278, 90, \dodoi{10.1126/science.278.5335.90}

\bibitem[{{Kama} {et~al.}(2019){Kama}, {Shorttle}, {Jermyn}, {Folsom}, {Furuya}, {Bergin}, {Walsh}, \& {Keller}}]{Kama19}
{Kama}, M., {Shorttle}, O., {Jermyn}, A.~S., {et~al.} 2019, \apj, 885, 114, \dodoi{10.3847/1538-4357/ab45f8}

\bibitem[{{Keyte} {et~al.}(2024{\natexlab{a}}){Keyte}, {Kama}, {Booth}, {Law}, \& {Leemker}}]{Keyte24_HD169142}
{Keyte}, L., {Kama}, M., {Booth}, A.~S., {Law}, C.~J., \& {Leemker}, M. 2024{\natexlab{a}}, \mnras, 534, 3576, \dodoi{10.1093/mnras/stae2314}

\bibitem[{{Keyte} {et~al.}(2024{\natexlab{b}}){Keyte}, {Kama}, {Chuang}, {Cleeves}, {Drozdovskaya}, {Furuya}, {Rawlings}, \& {Shorttle}}]{Keyte24}
{Keyte}, L., {Kama}, M., {Chuang}, K.-J., {et~al.} 2024{\natexlab{b}}, \mnras, 528, 388, \dodoi{10.1093/mnras/stae019}

\bibitem[{{Keyte} {et~al.}(2023){Keyte}, {Kama}, {Booth}, {Bergin}, {Cleeves}, {van Dishoeck}, {Drozdovskaya}, {Furuya}, {Rawlings}, {Shorttle}, \& {Walsh}}]{Keyte23}
{Keyte}, L., {Kama}, M., {Booth}, A.~S., {et~al.} 2023, Nature Astronomy, 7, 684, \dodoi{10.1038/s41550-023-01951-9}

\bibitem[{{Kim} \& {Yamamoto}(2003)}]{Kim03}
{Kim}, E., \& {Yamamoto}, S. 2003, Journal of Molecular Spectroscopy, 219, 296, \dodoi{10.1016/S0022-2852(03)00027-4}

\bibitem[{{Klaassen} {et~al.}(2013){Klaassen}, {Juhasz}, {Mathews}, {Mottram}, {De Gregorio-Monsalvo}, {van Dishoeck}, {Takahashi}, {Akiyama}, {Chapillon}, {Espada}, {Hales}, {Hogerheijde}, {Rawlings}, {Schmalzl}, \& {Testi}}]{Klaassen13}
{Klaassen}, P.~D., {Juhasz}, A., {Mathews}, G.~S., {et~al.} 2013, \aap, 555, A73, \dodoi{10.1051/0004-6361/201321129}

\bibitem[{{Laas} \& {Caselli}(2019)}]{Laas19}
{Laas}, J.~C., \& {Caselli}, P. 2019, \aap, 624, A108, \dodoi{10.1051/0004-6361/201834446}

\bibitem[{{Law} {et~al.}(2023){Law}, {Booth}, \& {{\"O}berg}}]{Law23_HD16}
{Law}, C.~J., {Booth}, A.~S., \& {{\"O}berg}, K.~I. 2023, \apjl, 952, L19, \dodoi{10.3847/2041-8213/acdfd0}

\bibitem[{{Law} {et~al.}(2021{\natexlab{a}}){Law}, {Loomis}, {Teague}, {{\"O}berg}, {Czekala}, {Andrews}, {Huang}, {Aikawa}, {Alarc{\'o}n}, {Bae}, {Bergin}, {Bergner}, {Boehler}, {Booth}, {Bosman}, {Calahan}, {Cataldi}, {Cleeves}, {Furuya}, {Guzm{\'a}n}, {Ilee}, {Le Gal}, {Liu}, {Long}, {M{\'e}nard}, {Nomura}, {Qi}, {Schwarz}, {Sierra}, {Tsukagoshi}, {Yamato}, {van't Hoff}, {Walsh}, {Wilner}, \& {Zhang}}]{LawMAPSIII}
{Law}, C.~J., {Loomis}, R.~A., {Teague}, R., {et~al.} 2021{\natexlab{a}}, \apjs, 257, 3, \dodoi{10.3847/1538-4365/ac1434}

\bibitem[{{Law} {et~al.}(2021{\natexlab{b}}){Law}, {Teague}, {Loomis}, {Bae}, {{\"O}berg}, {Czekala}, {Andrews}, {Aikawa}, {Alarc{\'o}n}, {Bergin}, {Bergner}, {Booth}, {Bosman}, {Calahan}, {Cataldi}, {Cleeves}, {Furuya}, {Guzm{\'a}n}, {Huang}, {Ilee}, {Le Gal}, {Liu}, {Long}, {M{\'e}nard}, {Nomura}, {P{\'e}rez}, {Qi}, {Schwarz}, {Soto}, {Tsukagoshi}, {Yamato}, {van't Hoff}, {Walsh}, {Wilner}, \& {Zhang}}]{Law21_MAPSIV}
{Law}, C.~J., {Teague}, R., {Loomis}, R.~A., {et~al.} 2021{\natexlab{b}}, \apjs, 257, 4, \dodoi{10.3847/1538-4365/ac1439}

\bibitem[{{Le Gal} {et~al.}(2019){Le Gal}, {{\"O}berg}, {Loomis}, {Pegues}, \& {Bergner}}]{Legal19}
{Le Gal}, R., {{\"O}berg}, K.~I., {Loomis}, R.~A., {Pegues}, J., \& {Bergner}, J.~B. 2019, \apj, 876, 72, \dodoi{10.3847/1538-4357/ab1416}

\bibitem[{{Le Gal} {et~al.}(2021){Le Gal}, {{\"O}berg}, {Teague}, {Loomis}, {Law}, {Walsh}, {Bergin}, {M{\'e}nard}, {Wilner}, {Andrews}, {Aikawa}, {Booth}, {Cataldi}, {Bergner}, {Bosman}, {Cleeves}, {Czekala}, {Furuya}, {Guzm{\'a}n}, {Huang}, {Ilee}, {Nomura}, {Qi}, {Schwarz}, {Tsukagoshi}, {Yamato}, \& {Zhang}}]{Legal21}
{Le Gal}, R., {{\"O}berg}, K.~I., {Teague}, R., {et~al.} 2021, \apjs, 257, 12, \dodoi{10.3847/1538-4365/ac2583}

\bibitem[{{Loomis} {et~al.}(2018){Loomis}, {Cleeves}, {{\"O}berg}, {Aikawa}, {Bergner}, {Furuya}, {Guzman}, \& {Walsh}}]{Loomis18}
{Loomis}, R.~A., {Cleeves}, L.~I., {{\"O}berg}, K.~I., {et~al.} 2018, \apj, 859, 131, \dodoi{10.3847/1538-4357/aac169}

\bibitem[{{Loomis} {et~al.}(2020){Loomis}, {{\"O}berg}, {Andrews}, {Bergin}, {Bergner}, {Blake}, {Cleeves}, {Czekala}, {Huang}, {Le Gal}, {M{\'e}nard}, {Pegues}, {Qi}, {Walsh}, {Williams}, \& {Wilner}}]{Loomis20}
{Loomis}, R.~A., {{\"O}berg}, K.~I., {Andrews}, S.~M., {et~al.} 2020, \apj, 893, 101, \dodoi{10.3847/1538-4357/ab7cc8}

\bibitem[{{Ma} {et~al.}(2024){Ma}, {Quan}, {Zhou}, {Esimbek}, {Li}, {Li}, {Zhang}, {Tuo}, \& {Feng}}]{Ma24}
{Ma}, R., {Quan}, D., {Zhou}, Y., {et~al.} 2024, Research in Astronomy and Astrophysics, 24, 075017, \dodoi{10.1088/1674-4527/ad5771}

\bibitem[{{McMullin} {et~al.}(2007){McMullin}, {Waters}, {Schiebel}, {Young}, \& {Golap}}]{McMullin_etal_2007}
{McMullin}, J.~P., {Waters}, B., {Schiebel}, D., {Young}, W., \& {Golap}, K. 2007, in Astronomical Society of the Pacific Conference Series, Vol. 376, Astronomical Data Analysis Software and Systems XVI, ed. R.~A. {Shaw}, F.~{Hill}, \& D.~J. {Bell}, 127

\bibitem[{{Millar} \& {Herbst}(1990)}]{Millar90}
{Millar}, T.~J., \& {Herbst}, E. 1990, \aap, 231, 466

\bibitem[{{Miotello} {et~al.}(2019){Miotello}, {Facchini}, {van Dishoeck}, {Cazzoletti}, {Testi}, {Williams}, {Ansdell}, {van Terwisga}, \& {van der Marel}}]{Miotello19}
{Miotello}, A., {Facchini}, S., {van Dishoeck}, E.~F., {et~al.} 2019, \aap, 631, A69, \dodoi{10.1051/0004-6361/201935441}

\bibitem[{{M{\"u}ller} {et~al.}(2005){M{\"u}ller}, {Schl{\"o}der}, {Stutzki}, \& {Winnewisser}}]{Muller05}
{M{\"u}ller}, H. S.~P., {Schl{\"o}der}, F., {Stutzki}, J., \& {Winnewisser}, G. 2005, Journal of Molecular Structure, 742, 215, \dodoi{10.1016/j.molstruc.2005.01.027}

\bibitem[{{M{\"u}ller} {et~al.}(2001){M{\"u}ller}, {Thorwirth}, {Roth}, \& {Winnewisser}}]{Muller01}
{M{\"u}ller}, H.~S.~P., {Thorwirth}, S., {Roth}, D.~A., \& {Winnewisser}, G. 2001, \aap, 370, L49, \dodoi{10.1051/0004-6361:20010367}

\bibitem[{{M{\"u}ller} {et~al.}(2019){M{\"u}ller}, {Maeda}, {Thorwirth}, {Lewen}, {Schlemmer}, {Medvedev}, {Winnewisser}, {De Lucia}, \& {Herbst}}]{Muller19}
{M{\"u}ller}, H. S.~P., {Maeda}, A., {Thorwirth}, S., {et~al.} 2019, \aap, 621, A143, \dodoi{10.1051/0004-6361/201834517}

\bibitem[{{Muro-Arena} {et~al.}(2018){Muro-Arena}, {Dominik}, {Waters}, {Min}, {Klarmann}, {Ginski}, {Isella}, {Benisty}, {Pohl}, {Garufi}, {Hagelberg}, {Langlois}, {Menard}, {Pinte}, {Sezestre}, {van der Plas}, {Villenave}, {Delboulb{\'e}}, {Magnard}, {M{\"o}ller-Nilsson}, {Pragt}, {Rabou}, \& {Roelfsema}}]{Muro_Arena18}
{Muro-Arena}, G.~A., {Dominik}, C., {Waters}, L.~B.~F.~M., {et~al.} 2018, \aap, 614, A24, \dodoi{10.1051/0004-6361/201732299}

\bibitem[{{{\"O}berg} {et~al.}(2021){{\"O}berg}, {Guzm{\'a}n}, {Walsh}, {Aikawa}, {Bergin}, {Law}, {Loomis}, {Alarc{\'o}n}, {Andrews}, {Bae}, {Bergner}, {Boehler}, {Booth}, {Bosman}, {Calahan}, {Cataldi}, {Cleeves}, {Czekala}, {Furuya}, {Huang}, {Ilee}, {Kurtovic}, {Le Gal}, {Liu}, {Long}, {M{\'e}nard}, {Nomura}, {P{\'e}rez}, {Qi}, {Schwarz}, {Sierra}, {Teague}, {Tsukagoshi}, {Yamato}, {van't Hoff}, {Waggoner}, {Wilner}, \& {Zhang}}]{Oberg21_MAPSI}
{{\"O}berg}, K.~I., {Guzm{\'a}n}, V.~V., {Walsh}, C., {et~al.} 2021, \apjs, 257, 1, \dodoi{10.3847/1538-4365/ac1432}

\bibitem[{{Pacheco-V{\'a}zquez} {et~al.}(2016){Pacheco-V{\'a}zquez}, {Fuente}, {Baruteau}, {Bern{\'e}}, {Ag{\'u}ndez}, {Neri}, {Goicoechea}, {Cernicharo}, \& {Bachiller}}]{Pacheco16}
{Pacheco-V{\'a}zquez}, S., {Fuente}, A., {Baruteau}, C., {et~al.} 2016, \aap, 589, A60, \dodoi{10.1051/0004-6361/201527089}

\bibitem[{{Paneque-Carre{\~n}o} {et~al.}(2024){Paneque-Carre{\~n}o}, {Izquierdo}, {Teague}, {Miotello}, {Bergin}, {Loomis}, \& {van Dishoeck}}]{Paneque24}
{Paneque-Carre{\~n}o}, T., {Izquierdo}, A.~F., {Teague}, R., {et~al.} 2024, \aap, 684, A174, \dodoi{10.1051/0004-6361/202347757}

\bibitem[{{Paneque-Carre{\~n}o} {et~al.}(2023){Paneque-Carre{\~n}o}, {Miotello}, {van Dishoeck}, {Tabone}, {Izquierdo}, \& {Facchini}}]{Paneque_MAPS}
{Paneque-Carre{\~n}o}, T., {Miotello}, A., {van Dishoeck}, E.~F., {et~al.} 2023, \aap, 669, A126, \dodoi{10.1051/0004-6361/202244428}

\bibitem[{{Paquette} {et~al.}(2017){Paquette}, {Hornung}, {Stenzel}, {Ryn{\"o}}, {Silen}, {Kissel}, \& {Hilchenbach}}]{Paquette17}
{Paquette}, J.~A., {Hornung}, K., {Stenzel}, O.~J., {et~al.} 2017, \mnras, 469, S230, \dodoi{10.1093/mnras/stx1623}

\bibitem[{{Phuong} {et~al.}(2018){Phuong}, {Chapillon}, {Majumdar}, {Dutrey}, {Guilloteau}, {Pi{\'e}tu}, {Wakelam}, {Diep}, {Tang}, {Beck}, \& {Bary}}]{Phuong18}
{Phuong}, N.~T., {Chapillon}, E., {Majumdar}, L., {et~al.} 2018, \aap, 616, L5, \dodoi{10.1051/0004-6361/201833766}

\bibitem[{{Phuong} {et~al.}(2021){Phuong}, {Dutrey}, {Chapillon}, {Guilloteau}, {Bary}, {Beck}, {Coutens}, {Denis-Alpizar}, {Di Folco}, {Diep}, {Majumdar}, {Melisse}, {Lee}, {Pietu}, {Stoecklin}, \& {Tang}}]{Phuong21}
{Phuong}, N.~T., {Dutrey}, A., {Chapillon}, E., {et~al.} 2021, \aap, 653, L5, \dodoi{10.1051/0004-6361/202141881}

\bibitem[{{Pinte} {et~al.}(2018{\natexlab{a}}){Pinte}, {Price}, {M{\'e}nard}, {Duch{\^e}ne}, {Dent}, {Hill}, {de Gregorio-Monsalvo}, {Hales}, \& {Mentiplay}}]{Pinte18_HD16}
{Pinte}, C., {Price}, D.~J., {M{\'e}nard}, F., {et~al.} 2018{\natexlab{a}}, \apjl, 860, L13, \dodoi{10.3847/2041-8213/aac6dc}

\bibitem[{{Pinte} {et~al.}(2018{\natexlab{b}}){Pinte}, {M{\'e}nard}, {Duch{\^e}ne}, {Hill}, {Dent}, {Woitke}, {Maret}, {van der Plas}, {Hales}, {Kamp}, {Thi}, {de Gregorio-Monsalvo}, {Rab}, {Quanz}, {Avenhaus}, {Carmona}, \& {Casassus}}]{pinte18}
{Pinte}, C., {M{\'e}nard}, F., {Duch{\^e}ne}, G., {et~al.} 2018{\natexlab{b}}, \aap, 609, A47, \dodoi{10.1051/0004-6361/201731377}

\bibitem[{{Pinte} {et~al.}(2020){Pinte}, {Price}, {M{\'e}nard}, {Duch{\^e}ne}, {Christiaens}, {Andrews}, {Huang}, {Hill}, {van der Plas}, {Perez}, {Isella}, {Boehler}, {Dent}, {Mentiplay}, \& {Loomis}}]{Pinte20}
{Pinte}, C., {Price}, D.~J., {M{\'e}nard}, F., {et~al.} 2020, \apjl, 890, L9, \dodoi{10.3847/2041-8213/ab6dda}

\bibitem[{{Price-Whelan} {et~al.}(2018){Price-Whelan}, {Sip{\H{o}}cz}, {G{\"u}nther}, {Lim}, {Crawford}, {Conseil}, {Shupe}, {Craig}, {Dencheva}, {Ginsburg}, {VanderPlas}, {Bradley}, {P{\'e}rez-Su{\'a}rez}, {de Val-Borro}, {Paper Contributors}, {Aldcroft}, {Cruz}, {Robitaille}, {Tollerud}, {Coordination Committee}, {Ardelean}, {Babej}, {Bach}, {Bachetti}, {Bakanov}, {Bamford}, {Barentsen}, {Barmby}, {Baumbach}, {Berry}, {Biscani}, {Boquien}, {Bostroem}, {Bouma}, {Brammer}, {Bray}, {Breytenbach}, {Buddelmeijer}, {Burke}, {Calderone}, {Cano Rodr{\'\i}guez}, {Cara}, {Cardoso}, {Cheedella}, {Copin}, {Corrales}, {Crichton}, {D{\textquoteright}Avella}, {Deil}, {Depagne}, {Dietrich}, {Donath}, {Droettboom}, {Earl}, {Erben}, {Fabbro}, {Ferreira}, {Finethy}, {Fox}, {Garrison}, {Gibbons}, {Goldstein}, {Gommers}, {Greco}, {Greenfield}, {Groener}, {Grollier}, {Hagen}, {Hirst}, {Homeier}, {Horton}, {Hosseinzadeh}, {Hu}, {Hunkeler}, {Ivezi{\'c}}, {Jain}, {Jenness}, {Kanarek}, {Kendrew}, {Kern}, {Kerzendorf}, {Khvalko},
  {King}, {Kirkby}, {Kulkarni}, {Kumar}, {Lee}, {Lenz}, {Littlefair}, {Ma}, {Macleod}, {Mastropietro}, {McCully}, {Montagnac}, {Morris}, {Mueller}, {Mumford}, {Muna}, {Murphy}, {Nelson}, {Nguyen}, {Ninan}, {N{\"o}the}, {Ogaz}, {Oh}, {Parejko}, {Parley}, {Pascual}, {Patil}, {Patil}, {Plunkett}, {Prochaska}, {Rastogi}, {Reddy Janga}, {Sabater}, {Sakurikar}, {Seifert}, {Sherbert}, {Sherwood-Taylor}, {Shih}, {Sick}, {Silbiger}, {Singanamalla}, {Singer}, {Sladen}, {Sooley}, {Sornarajah}, {Streicher}, {Teuben}, {Thomas}, {Tremblay}, {Turner}, {Terr{\'o}n}, {van Kerkwijk}, {de la Vega}, {Watkins}, {Weaver}, {Whitmore}, {Woillez}, {Zabalza}, \& {Contributors}}]{astropy_2018}
{Price-Whelan}, A.~M., {Sip{\H{o}}cz}, B.~M., {G{\"u}nther}, H.~M., {et~al.} 2018, \aj, 156, 123, \dodoi{10.3847/1538-3881/aabc4f}

\bibitem[{{Qi} {et~al.}(2011){Qi}, {D'Alessio}, {{\"O}berg}, {Wilner}, {Hughes}, {Andrews}, \& {Ayala}}]{Qi11}
{Qi}, C., {D'Alessio}, P., {{\"O}berg}, K.~I., {et~al.} 2011, \apj, 740, 84, \dodoi{10.1088/0004-637X/740/2/84}

\bibitem[{{Rab} {et~al.}(2020){Rab}, {Kamp}, {Dominik}, {Ginski}, {Muro-Arena}, {Thi}, {Waters}, \& {Woitke}}]{Rab20}
{Rab}, C., {Kamp}, I., {Dominik}, C., {et~al.} 2020, \aap, 642, A165, \dodoi{10.1051/0004-6361/202038712}

\bibitem[{{Rampinelli} {et~al.}(2024){Rampinelli}, {Facchini}, {Leemker}, {Bae}, {Benisty}, {Teague}, {Law}, {{\"O}berg}, {Portilla-Revelo}, \& {Cridland}}]{Rampinelli24}
{Rampinelli}, L., {Facchini}, S., {Leemker}, M., {et~al.} 2024, \aap, 689, A65, \dodoi{10.1051/0004-6361/202449698}

\bibitem[{{Ranjan} {et~al.}(2018){Ranjan}, {Todd}, {Sutherland}, \& {Sasselov}}]{Ranjan18}
{Ranjan}, S., {Todd}, Z.~R., {Sutherland}, J.~D., \& {Sasselov}, D.~D. 2018, Astrobiology, 18, 1023, \dodoi{10.1089/ast.2017.1770}

\bibitem[{{Rich} {et~al.}(2020){Rich}, {Wisniewski}, {Sitko}, {Grady}, {Tobin}, \& {Fukagawa}}]{Rich20_HST}
{Rich}, E.~A., {Wisniewski}, J.~P., {Sitko}, M.~L., {et~al.} 2020, \apj, 902, 4, \dodoi{10.3847/1538-4357/abb2a3}

\bibitem[{{Rivi{\`e}re-Marichalar} {et~al.}(2022){Rivi{\`e}re-Marichalar}, {Fuente}, {Esplugues}, {Wakelam}, {le Gal}, {Baruteau}, {Ribas}, {Mac{\'\i}as}, {Neri}, \& {Navarro-Almaida}}]{Marichalar22}
{Rivi{\`e}re-Marichalar}, P., {Fuente}, A., {Esplugues}, G., {et~al.} 2022, \aap, 665, A61, \dodoi{10.1051/0004-6361/202142906}

\bibitem[{{Rosenfeld} {et~al.}(2013){Rosenfeld}, {Andrews}, {Hughes}, {Wilner}, \& {Qi}}]{Rosenfeld13}
{Rosenfeld}, K.~A., {Andrews}, S.~M., {Hughes}, A.~M., {Wilner}, D.~J., \& {Qi}, C. 2013, \apj, 774, 16, \dodoi{10.1088/0004-637X/774/1/16}

\bibitem[{{Rosotti} {et~al.}(2021){Rosotti}, {Ilee}, {Facchini}, {Tazzari}, {Booth}, {Clarke}, \& {Kama}}]{Rosotti21}
{Rosotti}, G.~P., {Ilee}, J.~D., {Facchini}, S., {et~al.} 2021, \mnras, 501, 3427, \dodoi{10.1093/mnras/staa3869}

\bibitem[{{Ruffle} {et~al.}(1999){Ruffle}, {Hartquist}, {Caselli}, \& {Williams}}]{Ruffle99}
{Ruffle}, D.~P., {Hartquist}, T.~W., {Caselli}, P., \& {Williams}, D.~A. 1999, \mnras, 306, 691, \dodoi{10.1046/j.1365-8711.1999.02562.x}

\bibitem[{{Salinas} {et~al.}(2017){Salinas}, {Hogerheijde}, {Mathews}, {{\"O}berg}, {Qi}, {Williams}, \& {Wilner}}]{Salinas17}
{Salinas}, V.~N., {Hogerheijde}, M.~R., {Mathews}, G.~S., {et~al.} 2017, \aap, 606, A125, \dodoi{10.1051/0004-6361/201731223}

\bibitem[{{Semenov} {et~al.}(2018){Semenov}, {Favre}, {Fedele}, {Guilloteau}, {Teague}, {Henning}, {Dutrey}, {Chapillon}, {Hersant}, \& {Pi{\'e}tu}}]{Semenov18}
{Semenov}, D., {Favre}, C., {Fedele}, D., {et~al.} 2018, \aap, 617, A28, \dodoi{10.1051/0004-6361/201832980}

\bibitem[{{Shirley}(2015)}]{Shirley15}
{Shirley}, Y.~L. 2015, \pasp, 127, 299, \dodoi{10.1086/680342}

\bibitem[{{Sierra} {et~al.}(2021){Sierra}, {P{\'e}rez}, {Zhang}, {Law}, {Guzm{\'a}n}, {Qi}, {Bosman}, {{\"O}berg}, {Andrews}, {Long}, {Teague}, {Booth}, {Walsh}, {Wilner}, {M{\'e}nard}, {Cataldi}, {Czekala}, {Bae}, {Huang}, {Bergner}, {Ilee}, {Benisty}, {Le Gal}, {Loomis}, {Tsukagoshi}, {Liu}, {Yamato}, \& {Aikawa}}]{Sierra21_MAPS}
{Sierra}, A., {P{\'e}rez}, L.~M., {Zhang}, K., {et~al.} 2021, \apjs, 257, 14, \dodoi{10.3847/1538-4365/ac1431}

\bibitem[{{Speedie} {et~al.}(2025){Speedie}, {Dong}, {Teague}, {Segura-Cox}, {Pineda}, {Calcino}, {Longarini}, {Hall}, {Tang}, {Hashimoto}, {Paneque-Carre{\~n}o}, {Lodato}, \& {Veronesi}}]{Speedie25}
{Speedie}, J., {Dong}, R., {Teague}, R., {et~al.} 2025, arXiv e-prints, arXiv:2503.01957, \dodoi{10.48550/arXiv.2503.01957}

\bibitem[{{Teague}(2019)}]{Teague19JOSS}
{Teague}, R. 2019, The Journal of Open Source Software, 4, 1632, \dodoi{10.21105/joss.01632}

\bibitem[{Teague(2020)}]{rich_teague_2020_4321137}
Teague, R. 2020, richteague/keplerian\_mask: Initial Release, 1.0,  Zenodo, \dodoi{10.5281/zenodo.4321137}

\bibitem[{{Teague} {et~al.}(2019{\natexlab{a}}){Teague}, {Bae}, \& {Bergin}}]{Teague_19Natur}
{Teague}, R., {Bae}, J., \& {Bergin}, E.~A. 2019{\natexlab{a}}, \nat, 574, 378, \dodoi{10.1038/s41586-019-1642-0}

\bibitem[{{Teague} {et~al.}(2019{\natexlab{b}}){Teague}, {Bae}, \& {Bergin}}]{Teague19Natur}
---. 2019{\natexlab{b}}, \nat, 574, 378, \dodoi{10.1038/s41586-019-1642-0}

\bibitem[{{Teague} {et~al.}(2018{\natexlab{a}}){Teague}, {Bae}, {Bergin}, {Birnstiel}, \& {Foreman-Mackey}}]{Teague18_HD163296}
{Teague}, R., {Bae}, J., {Bergin}, E.~A., {Birnstiel}, T., \& {Foreman-Mackey}, D. 2018{\natexlab{a}}, \apjl, 860, L12, \dodoi{10.3847/2041-8213/aac6d7}

\bibitem[{{Teague} \& {Foreman-Mackey}(2018)}]{Teague18_bettermoments}
{Teague}, R., \& {Foreman-Mackey}, D. 2018, {Bettermoments: A Robust Method To Measure Line Centroids}, v1.0,  Zenodo, \dodoi{10.5281/zenodo.1419754}

\bibitem[{Teague {et~al.}(2021)Teague, Law, Huang, \& Meng}]{disksurf_Teague}
Teague, R., Law, C.~J., Huang, J., \& Meng, F. 2021, Journal of Open Source Software, 6, 3827, \dodoi{10.21105/joss.03827}

\bibitem[{{Teague} {et~al.}(2016){Teague}, {Guilloteau}, {Semenov}, {Henning}, {Dutrey}, {Pi{\'e}tu}, {Birnstiel}, {Chapillon}, {Hollenbach}, \& {Gorti}}]{Teague16}
{Teague}, R., {Guilloteau}, S., {Semenov}, D., {et~al.} 2016, \aap, 592, A49, \dodoi{10.1051/0004-6361/201628550}

\bibitem[{{Teague} {et~al.}(2018{\natexlab{b}}){Teague}, {Henning}, {Guilloteau}, {Bergin}, {Semenov}, {Dutrey}, {Flock}, {Gorti}, \& {Birnstiel}}]{Teague18_CS}
{Teague}, R., {Henning}, T., {Guilloteau}, S., {et~al.} 2018{\natexlab{b}}, \apj, 864, 133, \dodoi{10.3847/1538-4357/aad80e}

\bibitem[{{Teague} {et~al.}(2021){Teague}, {Bae}, {Aikawa}, {Andrews}, {Bergin}, {Bergner}, {Boehler}, {Booth}, {Bosman}, {Cataldi}, {Czekala}, {Guzm{\'a}n}, {Huang}, {Ilee}, {Law}, {Le Gal}, {Long}, {Loomis}, {M{\'e}nard}, {{\"O}berg}, {P{\'e}rez}, {Schwarz}, {Sierra}, {Walsh}, {Wilner}, {Yamato}, \& {Zhang}}]{Teague21}
{Teague}, R., {Bae}, J., {Aikawa}, Y., {et~al.} 2021, \apjs, 257, 18, \dodoi{10.3847/1538-4365/ac1438}

\bibitem[{{Teague} {et~al.}(2022){Teague}, {Bae}, {Andrews}, {Benisty}, {Bergin}, {Facchini}, {Huang}, {Longarini}, \& {Wilner}}]{Teague22}
{Teague}, R., {Bae}, J., {Andrews}, S.~M., {et~al.} 2022, \apj, 936, 163, \dodoi{10.3847/1538-4357/ac88ca}

\bibitem[{{Temmink} {et~al.}(2023){Temmink}, {Booth}, {van der Marel}, \& {van Dishoeck}}]{Temmink23}
{Temmink}, M., {Booth}, A.~S., {van der Marel}, N., \& {van Dishoeck}, E.~F. 2023, \aap, 675, A131, \dodoi{10.1051/0004-6361/202346272}

\bibitem[{{Thi} {et~al.}(2004){Thi}, {van Zadelhoff}, \& {van Dishoeck}}]{Thi04}
{Thi}, W.~F., {van Zadelhoff}, G.~J., \& {van Dishoeck}, E.~F. 2004, \aap, 425, 955, \dodoi{10.1051/0004-6361:200400026}

\bibitem[{{van der Marel} {et~al.}(2013){van der Marel}, {van Dishoeck}, {Bruderer}, {Birnstiel}, {Pinilla}, {Dullemond}, {van Kempen}, {Schmalzl}, {Brown}, {Herczeg}, {Mathews}, \& {Geers}}]{vanderMarel13}
{van der Marel}, N., {van Dishoeck}, E.~F., {Bruderer}, S., {et~al.} 2013, Science, 340, 1199, \dodoi{10.1126/science.1236770}

\bibitem[{{van der Plas} {et~al.}(2014){van der Plas}, {Casassus}, {M{\'e}nard}, {Perez}, {Thi}, {Pinte}, \& {Christiaens}}]{vanderPlas14}
{van der Plas}, G., {Casassus}, S., {M{\'e}nard}, F., {et~al.} 2014, \apjl, 792, L25, \dodoi{10.1088/2041-8205/792/2/L25}

\bibitem[{{van der Tak} {et~al.}(2007){van der Tak}, {Black}, {Sch{\"o}ier}, {Jansen}, \& {van Dishoeck}}]{vanderTak07}
{van der Tak}, F.~F.~S., {Black}, J.~H., {Sch{\"o}ier}, F.~L., {Jansen}, D.~J., \& {van Dishoeck}, E.~F. 2007, \aap, 468, 627, \dodoi{10.1051/0004-6361:20066820}

\bibitem[{{van der Velden}(2020)}]{vanderVelden20}
{van der Velden}, E. 2020, The Journal of Open Source Software, 5, 2004, \dodoi{10.21105/joss.02004}

\bibitem[{{van der Walt} {et~al.}(2011){van der Walt}, {Colbert}, \& {Varoquaux}}]{vanderWalt_etal_2011}
{van der Walt}, S., {Colbert}, S.~C., \& {Varoquaux}, G. 2011, Computing in Science and Engineering, 13, 22, \dodoi{10.1109/MCSE.2011.37}

\bibitem[{{Vastel} {et~al.}(2018){Vastel}, {Qu{\'e}nard}, {Le Gal}, {Wakelam}, {Andrianasolo}, {Caselli}, {Vidal}, {Ceccarelli}, {Lefloch}, \& {Bachiller}}]{Vastel18}
{Vastel}, C., {Qu{\'e}nard}, D., {Le Gal}, R., {et~al.} 2018, \mnras, 478, 5514, \dodoi{10.1093/mnras/sty1336}

\bibitem[{{Visser} {et~al.}(2018){Visser}, {Bruderer}, {Cazzoletti}, {Facchini}, {Heays}, \& {van Dishoeck}}]{Visser18}
{Visser}, R., {Bruderer}, S., {Cazzoletti}, P., {et~al.} 2018, \aap, 615, A75, \dodoi{10.1051/0004-6361/201731898}

\bibitem[{{Wichittanakom} {et~al.}(2020){Wichittanakom}, {Oudmaijer}, {Fairlamb}, {Mendigut{\'\i}a}, {Vioque}, \& {Ababakr}}]{Wichittanakom20}
{Wichittanakom}, C., {Oudmaijer}, R.~D., {Fairlamb}, J.~R., {et~al.} 2020, \mnras, 493, 234, \dodoi{10.1093/mnras/staa169}

\bibitem[{{Wilson}(1999)}]{Wilson99}
{Wilson}, T.~L. 1999, Reports on Progress in Physics, 62, 143, \dodoi{10.1088/0034-4885/62/2/002}

\bibitem[{{Xie} {et~al.}(2021){Xie}, {Haffert}, {de Boer}, {Kenworthy}, {Brinchmann}, {Girard}, {Snellen}, \& {Keller}}]{Xie21}
{Xie}, C., {Haffert}, S.~Y., {de Boer}, J., {et~al.} 2021, \aap, 650, L6, \dodoi{10.1051/0004-6361/202140602}

\bibitem[{{Yamato} {et~al.}(2025, in prep.){Yamato}, {Aikawa}, \& {Law}}]{Yamato25}
{Yamato}, Y., {Aikawa}, Y., \& {Law}, C.~J. 2025, in prep., in prep

\bibitem[{{Yoshida} {et~al.}(2024){Yoshida}, {Nomura}, {Law}, {Teague}, {Shibaike}, {Furuya}, \& {Tsukagoshi}}]{Yoshida24}
{Yoshida}, T.~C., {Nomura}, H., {Law}, C.~J., {et~al.} 2024, \apjl, 971, L15, \dodoi{10.3847/2041-8213/ad654c}

\bibitem[{{Zhang} {et~al.}(2021){Zhang}, {Booth}, {Law}, {Bosman}, {Schwarz}, {Bergin}, {{\"O}berg}, {Andrews}, {Guzm{\'a}n}, {Walsh}, {Qi}, {van't Hoff}, {Long}, {Wilner}, {Huang}, {Czekala}, {Ilee}, {Cataldi}, {Bergner}, {Aikawa}, {Teague}, {Bae}, {Loomis}, {Calahan}, {Alarc{\'o}n}, {M{\'e}nard}, {Le Gal}, {Sierra}, {Yamato}, {Nomura}, {Tsukagoshi}, {P{\'e}rez}, {Trapman}, {Liu}, \& {Furuya}}]{Zhang21}
{Zhang}, K., {Booth}, A.~S., {Law}, C.~J., {et~al.} 2021, \apjs, 257, 5, \dodoi{10.3847/1538-4365/ac1580}

\bibitem[{{Zhang} {et~al.}(2018){Zhang}, {Zhu}, {Huang}, {Guzm{\'a}n}, {Andrews}, {Birnstiel}, {Dullemond}, {Carpenter}, {Isella}, {P{\'e}rez}, {Benisty}, {Wilner}, {Baruteau}, {Bai}, \& {Ricci}}]{Zhang18}
{Zhang}, S., {Zhu}, Z., {Huang}, J., {et~al.} 2018, \apjl, 869, L47, \dodoi{10.3847/2041-8213/aaf744}

\end{thebibliography}
\bibliographystyle{aasjournal}



\end{document}